\documentclass{article}

\usepackage{amsmath,amssymb,amsthm}
\usepackage{a4wide}
\usepackage{subfig}
\usepackage{pdfsync}
\usepackage{graphicx}
\usepackage{epstopdf}
\usepackage{url}
\usepackage{morefloats}
\usepackage[usenames,dvipsnames]{pstricks}
\usepackage{epsfig}
\usepackage{pst-grad} 
\usepackage{pst-plot} 
\usepackage{siunitx}
\usepackage{enumitem}
\usepackage{booktabs}
\usepackage[superscript,biblabel]{cite}
\usepackage{hyperref}
\usepackage{etoolbox}
\appto\UrlBreaks{\do\-}

\renewcommand{\figurename}{\textbf{Fig.}}

\newtheorem*{remark}{Remark}

\renewcommand{\geq}{\geqslant}
\renewcommand{\leq}{\leqslant}

\begin{document}
	
\title{Optimal control of the COVID-19 pandemic:\\ 
controlled sanitary deconfinement in Portugal\thanks{Paper 
whose final and definite form is published Open Access by 
\emph{Scientific Reports}, ISSN 2045-2322, Springer Nature. 
Submitted 01/Sept/2020; Revised 16/Dec/2020; Accepted 27/Jan/2021;
Published by {\tt www.nature.com/scientificreports} [see 
{\tt https://doi.org/10.1038/s41598-021-83075-6}].}}

\author{\href{https://orcid.org/0000-0002-7238-546X}{Cristiana J. Silva$^{*,1}$}\\ 
\texttt{\footnotesize cjoaosilva@ua.pt} 
\and 
\href{https://orcid.org/0000-0003-4082-7523}{Carla Cruz$^{1}$}\\
\texttt{\footnotesize carla.cruz@ua.pt} 
\and 
\href{https://orcid.org/0000-0001-8641-2505}{Delfim F. M. Torres$^{1}$}\\
\texttt{\footnotesize delfim@ua.pt}
\and 
\href{https://orcid.org/0000-0002-0579-9347}{Alberto P. Mu\~nuzuri$^{2}$}\\
\texttt{\footnotesize alberto.perez.munuzuri@usc.es} 
\and
\href{https://orcid.org/0000-0002-0858-8545}{Alejandro Carballosa$^{2}$}\\
\texttt{\footnotesize ac.carballosa@gmail.com} 
\and
\href{https://orcid.org/0000-0003-0872-5017}{Iv\'{a}n Area$^{3}$}\\
\texttt{\footnotesize area@uvigo.gal} 
\and
\href{https://orcid.org/0000-0001-8202-6578}{Juan J. Nieto$^{4}$}\\
\texttt{\footnotesize juanjose.nieto.roig@usc.es}
\and
\href{https://orcid.org/0000-0001-6774-5363}{Rui Fonseca-Pinto$^{5}$}\\
\texttt{\footnotesize rui.pinto@ipleiria.pt} 
\and
\href{https://orcid.org/0000-0002-7766-576X}{Rui Passadouro$^{5,6}$}\\
\texttt{\footnotesize rmfonseca@arscentro.min-saude.pt} 
\and
\href{https://orcid.org/0000-0001-6567-1487}{Estev\~{a}o Soares dos Santos$^{6}$}\\
\texttt{\footnotesize essantos3@arscentro.min-saude.pt} 
\and
\href{https://orcid.org/0000-0002-0847-824X}{Wilson Abreu$^{7}$}\\
\texttt{\footnotesize wjabreu@esenf.pt} 
\and 
\href{https://orcid.org/0000-0002-6024-6294}{Jorge Mira$^{*,8}$}\\
\texttt{\footnotesize jorge.mira@usc.es}} 


\date{\small 
$^1$Center for Research and Development in Mathematics and Applications (CIDMA),\\
Department of Mathematics, University of Aveiro, 3810-193 Aveiro, Portugal\\[0.05cm]
$^2$Institute CRETUS, Group of Nonlinear Physics, Department of Physics,\\ 
Universidade de Santiago de Compostela, 15782 Santiago de Compostela, Spain\\[0.05cm]
$^3$Departamento de Matem\'atica Aplicada II, 
E. E. Aeron\'autica e do Espazo, Campus de Ourense, 
Universidade de Vigo, 32004 Ourense, Spain\\[0.05cm]
$^4$\text{Instituto de Matem\'aticas, Universidade de Santiago de Compostela,
15782 Santiago de Compostela, Spain}\\[0.05cm]
$^5$Center for Innovative Care and Health Technology (ciTechCare),
Polytechnic of Leiria, Portugal\\[0.10cm]
$^6$ACES Pinhal Litoral -- ARS Centro, Portugal\\[0.05cm]
$^7$School of Nursing \& Research Centre 
``Centre for Health Technology and Services Research / ESEP-CINTESIS'', Porto, Portugal\\[0.05cm]
$^8$Departamento de F\'{\i}sica Aplicada, Universidade de Santiago de Compostela,\\ 
15782 Santiago de Compostela, Spain}

\maketitle

	
\begin{abstract}
The COVID-19 pandemic has forced policy makers to decree urgent confinements 
to stop a rapid and massive contagion. However, after that stage, societies 
are being forced to find an equilibrium between the need to reduce contagion 
rates and the need to reopen their economies. The experience hitherto lived 
has provided data on the evolution of the pandemic, in particular the population 
dynamics as a result of the public health measures enacted.  This allows 
the formulation of forecasting mathematical models to anticipate the consequences 
of political decisions. Here we propose a model to do so and apply 
it to the case of Portugal. With a mathematical deterministic model, 
described by a system of ordinary differential equations, we fit 
the real evolution of COVID-19 in this country.
After identification of the population readiness to follow social 
restrictions, by analyzing the social media, we incorporate this effect in a version 
of the model that allow us to check different scenarios. This is realized by considering 
a Monte Carlo discrete version of the previous model coupled via a complex network. 
Then, we apply optimal control theory to maximize the number of people returning 
to ``normal life'' and minimizing the number of active infected individuals
with minimal economical costs while warranting a low level of hospitalizations. 
This work allows testing various scenarios of pandemic management 
(closure of sectors of the economy, partial/total compliance with protection 
measures by citizens, number of beds in intensive care units, etc.), 
ensuring the responsiveness of the health system, 
thus being a public health decision support tool. 
\end{abstract}	

	
\section*{}

COVID‑19 is an ongoing global concern.
On March 11, 2020, the World Health Organization (WHO) declared the state of pandemic 
due to SARS-COV2 infection and, worldwide, the containment strategies to control 
the spread of COVID-19 were gradually intensified. In the first three months after COVID-19 
emerged, nearly 1 million people were infected and 50,000 died. Although we had in the past 
similar diseases caused by the same family of virus (e.g., SARS and MERS), these strategies 
are still of huge importance as the rate of spread of the SARS-COV2 virus is higher 
\cite{Noah}. The social and clinical experience with COVID-19 will leave lasting marks
in society and in the health system, from Latin cultural habits
(proximity, touch, kiss) until health system configuration changes,
leaving hospitals for more complex clinical situations
and providing community institutions (Health Centers, Family Health Units 
and Integrated Continuous Care Units) with diagnostic and therapeutic means 
that avoid systematic recourse to hospital emergencies.

By August 15, 2020, the cumulated number of confirmed cases by COVID-19 
was of 21,387,974, with 14,169,695 recovered cases and 764,112 deaths, 
corresponding to 6,454,140 active cases (at a given time $t$, 
the term ``active infected'' corresponds to the number of confirmed infected 
individuals active at that time $t$, while the term ``confirmed infected'' 
corresponds to the accumulated number of confirmed infected individuals 
from the beginning of the epidemic till time $t$).
Regarding the active cases,  6,035,791 (99\%) suffer mild condition of the disease and 65,488 (1\%) 
are in serious or critical health situation \cite{worldometers}.
In Portugal, the first confirmed 2 infected cases were reported on March 2, 2020, 
and the Government ordered public services to draw up a contingency plan in line 
with the guidelines set by the Portuguese Public Health Authorities. On March 12, 2020, 
it was declared State of Emergency. In the following week, additional measures were adopted, 
such as: prohibition of events, meetings or gathering of people, regardless of reason or nature, 
with 100 or more people; prohibition of drinking alcoholic beverages in public open-air spaces, 
except for outdoor areas catering and beverage establishments, duly licensed for the purpose; 
documentary control of people in borders; the suspension of all and any activity of stomatology 
and dentistry, with the exception of proven urgent situations and non-postponable. 
Teaching as well as non-teaching and classroom training activities were suspended 
from 16th March 2020; \cite{aulas:susp}
the air traffic to and from Portugal was banned for all flights to and from countries 
that do not belong to the European Union, with certain exceptions. 
Actually, the Por­tuguese were advised to stay at home, avoiding social contacts, 
since 14th March 2020, inclusive, restricting to the maximum their exits from home.
From March 20 on, it was mandatory to adopt the teleworking regime, regardless of the employment relationship, 
whenever the functions in question allow. On May 2 the emergency status was canceled 
(duration of 45 days). After the 45 days of state of emergency, the Government progressively 
established measures for the reopening of the economy but with rules for the control 
of the spread of the virus. Portugal is still in situation of alert, and the situation 
of calamity and contingency can be declared, depending on the region and the number 
of active cases. According to the Portuguese Health Authorities, as of the writing, 
there has not been an overload of intensive care services; since the beginning of the Portuguese 
outbreak the intensive medicine capacity increased from 629 to 819 beds (+23\%) (data from June 14, 2020); 
the health authorities objective is to reach, by the end of 2020, a ratio of 9.4 beds 
per 100 thousand inhabitants. Moreover, Portugal did not enter a rupture situation; 
at the peak of the epidemic (in the end of April, beginning of May), 
there were 1026 intensive care beds; the levels of intensive medicine occupancy, 
by June 14, 2020, were of 61\% at national level and 65\% in the Lisbon 
and Vale do Tejo region \cite{cap:UCI:pt}.

The way we manage today the pandemic is related to the ability to produce quality data, 
which in turn will allow us to use the same data for mathematical modeling tasks, 
that are the best framework to deal with upcoming scenarios \cite{Metcalf}. 
Many efforts have been done in this field 
\cite{Giordano:modcovid:NatMed,Lopez:HumBehav2020,Hoertel:modFrance:Natmed,Kissler860}.
The adjustment of the model parameters in a dynamic way, 
through the imposition of limits on the system 
in order to optimize a given function, can be implemented through 
the theory of optimal control \cite{MR4091761}. 

The usefulness of optimal control in epidemiology is well-known:
while mathematical modeling of infectious diseases has shown that 
combinations of isolation, quarantine, vaccination and/or treatment are often 
necessary in order to eliminate an infectious disease, optimal control theory
tell us how they should be administered, by providing the right times 
for intervention and the right amounts \cite{MR3918295,MR3629459}.  
This optimization strategy has also been 
used in some works within the scope of \text{COVID-19}.
Optimal control of an adapted Susceptible--Exposure--Infection--Recovery 
(SEIR) model has been done with the aim to investigate the efficacy 
of two potential lockdown release strategies on the UK population \cite{Rawson:OC}.  
Other COVID-19 case studies include the use of optimal control in USA \cite{Tsay:OC:US}. 
Optimal administration of an hypothetical vaccine for COVID-19 
has been also investigated \cite{Libotte}; and an expression for the basic 
reproduction number in terms of the control variables obtained \cite{Lemecha}. 
According to the most recent pandemic spreading data, until a large immunization rate is achieved 
(ideally by a vaccine), the application of so-called nonpharmaceutical interventions (NPIs) 
is the key to control the number of active infected individuals.\cite{Zine}

Here we are interested in using optimal control theory has a tool 
to understand ways to curtail the spread of COVID-19 in Portugal 
by devising optimal disease intervention strategies. 
Moreover, we take into account several important
issues that have not yet been fully considered in the literature. 
Our model allows the application of the theory of optimal control, to test containment scenarios 
in which the response capacity of health services is maintained. 
Because the pandemic has shown that the public health concern is not only a medical problem, 
but also affects society as a whole \cite{Moradian:2020}, the dynamics 
of monitoring the containment measures, that allow each individual to remain 
in the protected $P$ class, is here obtained through models of analysis of social networks, 
which differentiates this study getting closer to the real behavior of individuals
and also predicting the adherence of the population to possible government policies.


\section*{Results}

\subsection*{Confirmed active infected individuals in Portugal}
\label{sec:2}

We propose a deterministic $SAIRP$ mathematical model for the transmission dynamics 
of SARS-CoV-2 in a homogeneous population, which is subdivided into five compartments 
depending on the state of infection and disease of the individuals 
(see Supplementary Fig.~1): $S$, susceptible (uninfected and not immune); 
$A$, infected but asymptomatic (undetected); $I$, 
active infected (symptomatic and detected/confirmed); $R$, removed 
(recovered and deaths by COVID-19); $P$, protected/prevented
(not infected, not immune, but that are under protective measures).  

The class $P$ represents all individuals that practice, with daily 
efficacy, the so-called non-pharmaceutical interventions (NPIs), e.g., 
physical distancing, use of face masks, and eye protection to prevent 
person-to-person transmission of SARS-CoV-2 and COVID-19. 
Based on recent literature,\cite{protective:Lancet,Haug:NatHB2020}
we assume that the individuals in the class $P$ are free from infection, 
but are not immune and, if they stop taking these measures, 
they become susceptible again, at a rate $\omega = w m$, 
where $w$ represents the transition rate from protected $P$ to susceptible $S$ and $m$ 
represents the fraction of protected individuals that is transferred from $P$ to $S$ class  
(see Supplementary Fig.~1 for the diagram of the model; 
for the equations and a description of the parameters, see the Methods section). 

\begin{figure}[!htb]
\centering 
\includegraphics[scale=0.42]{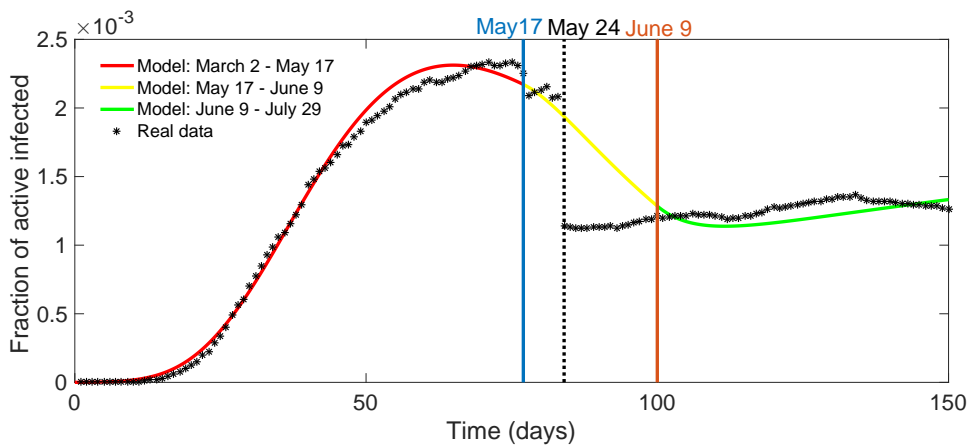}
\caption{\textbf{Fraction of confirmed active cases per day in Portugal.} 
Red line: from  March 2 to May 17, 2020. 
Yellow line: from May 17 to June 9, 2020. 
Green line: from June 9 to July 29, 2020. 
The drastic jump down in the real data (black points) corresponds to the day 
when the Portuguese authorities announced 9844 recovered individuals on May 24.}
\label{fig:Inf}
\end{figure}

In Fig.~\ref{fig:Inf}, we show that the SAIRP model 
(as described above and in detail in Methods) 
fits well the confirmed active infected cases in Portugal from March 2, 2020 
until July 29, 2020 (a total of 150 days), using the data from The Portuguese 
Public Health Authorities \cite{dgs-covid}. More precisely, based on daily reports 
from the Portuguese Public Health Authorities, that provide information 
about the confirmed infected cases, recovered, and deaths, the active cases 
are therefore the result of subtracting to the cumulative confirmed cases 
the sum of the recovered and deaths by COVID-19. See Section Methods 
for the parameter values and initial conditions used, as well as their justification. 

Most of the parameter values of the $SAIRP$ model are fixed for the 150 days considered. 
However, we analyzed the model in three different time intervals from the first confirmed 
case, on March 2, until July 29, and the parameters $\beta$, $p$ and $m$ take different 
values in these three time intervals. At first, we consider the time interval going 
from the first confirmed infected individual (March 2) until May 17, that is, 
15 days after the end of the three Emergency States in Portugal. 
Here, despite the fraction of susceptible individuals $S$ 
that are transferred to class $P$ being $p_1=0.675$ (see Table~\ref{table:param:values:PT} 
in Methods), meaning that approximately $67,5\%$ of the population 
was \emph{protected} due to the COVID-19 confinement policies 
during the three emergency states  
(suspension of activities in schools and universities, high risk groups 
protection and teleworking regime adoption) \cite{dgs-covid,legislacao:covid19},
the number of infected individuals  increased exponentially
(red curve in Fig.~\ref{fig:Inf}).
The second time interval goes from May 17 until June 9, 
the period when the number of new infected individuals 
grows slower comparing with the beginning of the outbreak. 
In this time period, and after the end of the three emergency states (during 45 days), 
the fraction of susceptible individuals that could stay \emph{protected} 
decreased ($p_2=0.55$), which, together with a low rate of $\beta_2=0.55$, 
explains the progressive decrease of $I$ (yellow curve in Fig.~\ref{fig:Inf}). 
Finally, the model was applied to the period going from June 9 until July 29, 2020. 
In that case, with the gradual opening of the society and economy, 
the value for $p_3$ becomes smaller and $\beta_3$ increases as the number 
of active infected individuals started to rise again
(green curve in Fig.~\ref{fig:Inf}).  
For these parameter values $\beta_i$, $p_i$, with $i = 1, 2, 3$, 
we estimated the parameter values $m_i$ (see Methods
for details on the estimation of the parameters).


\subsection*{Social opinion biased SAIRP model}
\label{sec:social}

The pandemic evolutions along past months, in different regions worldwide, 
demonstrated that the behavior of the population is of crucial influence. 
Same control policies, implemented in different regions, resulted in 
different outcomes. Even more, the same policies, implemented at different 
times, may produce different outcomes as the social state of opinion also changes with time.

We aim to incorporate the state of people's opinion 
into the SAIRP model in order to analyze its influence. The process 
is divided into three steps. First, we calculate, from empirical data, 
the social network describing the social interactions for Portugal 
at two different moments of time (April and July 2020). With this information, 
we consider a simple opinion model that provides a probability distribution 
function that we interpret as the distribution of opinions to follow government 
policies (distributed from zero to one, zero meaning no intention to accept 
the policies and one total acceptance). As a final step, we introduce this 
probability distribution function into the SAIRP model 
by modulating the access to class $P$.


\subsection*{Social opinion distributions}
\label{sec:social:opinion}

The details on the construction of the network, describing the social 
interactions, are explained in the Methods section. 
Just note that in both cases analyzed (April and July 2020) the network topology 
is quite different, reflecting a different social state. Each network is composed 
by a set of nodes (corresponding to different users or persons) and the connections 
with other nodes in the network. Both networks built, as described, constitute some 
kind of fingerprint of the social situation in Portugal 
at the specific periods of time considered.

\begin{figure}[!htb]
\centering 
\includegraphics[scale=1.0]{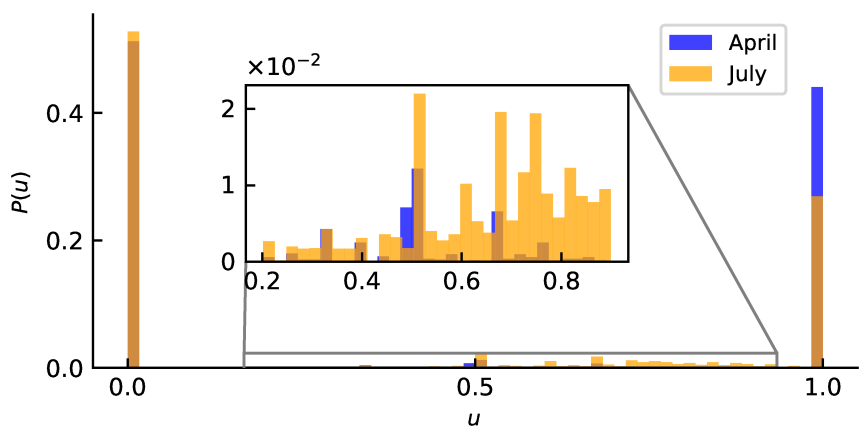}
\caption{\textbf{Probability distribution ($\mathbf{P(u)}$) for each opinion ($\mathbf{u}$).} 
The opinion ranges from zero to one, zero meaning no intention 
to follow the government policies while one means complete adhesion 
to this policy. The blue values correspond to the Portuguese situation 
in April 2020 while the yellow ones are for the situation in July 2020.}
\label{fig:opiniondistrib}
\end{figure}

We use this network topology in order to incorporate a model of opinion. 
For that, we consider now that each node in our network is endowed with 
some dynamical equations, which allow to determine its state of opinion, 
combined with the information that it is coming through the network. 
The opinion dynamical equations are based on the logistic equations 
and they are fully described in the Methods section. 
The combined effect of the opinion model for each node, together 
with the influence of the information coming through the network, 
results in an opinion distribution function. The results are presented 
in Fig.~\ref{fig:opiniondistrib}. To each opinion 
in the $x$-axis it corresponds a probability to occur. In the two cases 
considered (April and July 2020) the opinion distribution appears 
very polarized, but in July we can detect a clear decrease 
in the intention to follow government imposed policies. This reflects 
the experience of the situation as it happened, during the worst 
of the pandemic (April) people were eager to follow any policy 
that helped reducing the impact of the disease, while in July 
more people changed the opinion and decide to oppose the restriction policies.


\subsubsection*{SAIRP model with opinion distribution}
\label{sec:social:model}

Our aim now is to couple the previous SAIRP model with opinion distributions. 
For this purpose, instead of using a deterministic approach, we find more feasible 
a multi-agent based approach with stochastic dynamics, where a large number 
of individuals conform a mobility network and infected nodes can spread 
the disease through its connections with susceptible individuals 
\cite{Hoertel:modFrance:Natmed}. The considered 
synthetic population is built according to the Watts--Strogatz model \cite{Watts1998}, 
so it has small-world properties and high clustering. In particular, we considered 
a synthetic network with an average connectivity $\langle k \rangle =5$ and a probability 
of long range connections of $5\%$.  Following the main idea of the SAIRP model, 
each node can be in one of the different compartments. Susceptible nodes can become 
asymptomatic by interactions with either asymptomatic or infected nodes, or become 
protected with probability $\phi p$, at each time step. At the same time, asymptomatic 
individuals are detected with probability $\nu$ and confirmed infected individuals 
can recover with probability $\mu$. Finally, protected individuals become susceptible 
again with probability $\omega$. The network is initialized with a discrete number 
of infected individuals and then these processes are evaluated until the dynamics 
of the disease become stationary. 

We now introduce the opinion distributions through the protected $P$ compartment. 
Considering the opinion probability distributions, $P(u)$ (Fig.~\ref{fig:opiniondistrib}) 
for each node of the synthetic population we assign an opinion value drawn from $P(u)$. 
Next, instead of having a fixed value for $p$ and $m$, we consider that each node 
has its own probabilities of becoming protected and susceptible again, $p_i$ and $m_i$, 
and that these probabilities are given by the opinion value of the particular node. 
While we can directly identify $p_i$ with $u_i$, $m_i$ has to be related to the 
complementary of $u_i$: $\overline{u_i}=1-u_i$. Note that the meaning of the 
extreme values of the opinions are either to follow the directives and stay at home 
(if $u_i=1.0$) or not (if $u_i=0.0$). In this way, the opinion distributions 
overlap smoothly with the transition to the protected compartment. Finally, following 
the infection rate of the deterministic model, $\beta\cdot(1-p)$, we consider 
that the infection process occurs along the connection of an infected node $i$ 
with a susceptible node $j$ with probability $\beta\cdot (1-p_j)$. In this way, 
the infection process is also weighted by the opinion value of the susceptible node. 

\begin{remark}
Although the values of $p_j$ are directly related 
to the $u_j$ values, their index $j$ belong to completely different networks. 
On one hand, from the social network we extract the opinion distribution $P(u)$, 
from which we build a new distribution $P(p)$ with identical probabilities 
but applied to the epidemiological network (the one where we simulate 
the infective stochastic dynamics), assigning each node a value $p_j$.
\end{remark}

\begin{figure}[!htb]
\centering 
\subfloat[]{\label{fig:opinion_expersVSApril}
\includegraphics[scale=1.0]{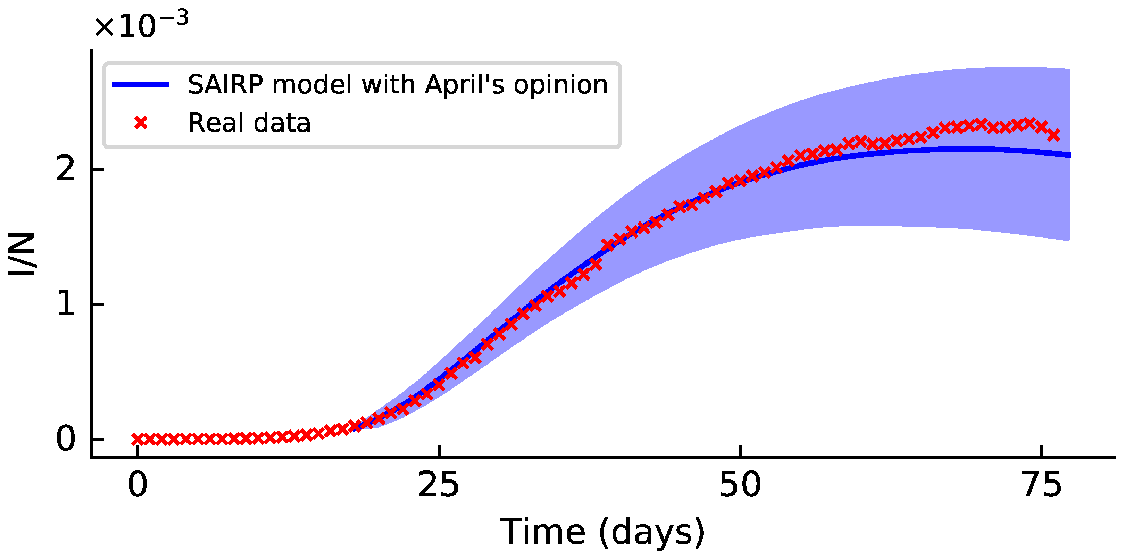}}\\
\subfloat[]{\label{fig:opinion_AprilVSJuly}
\includegraphics[scale=1.0]{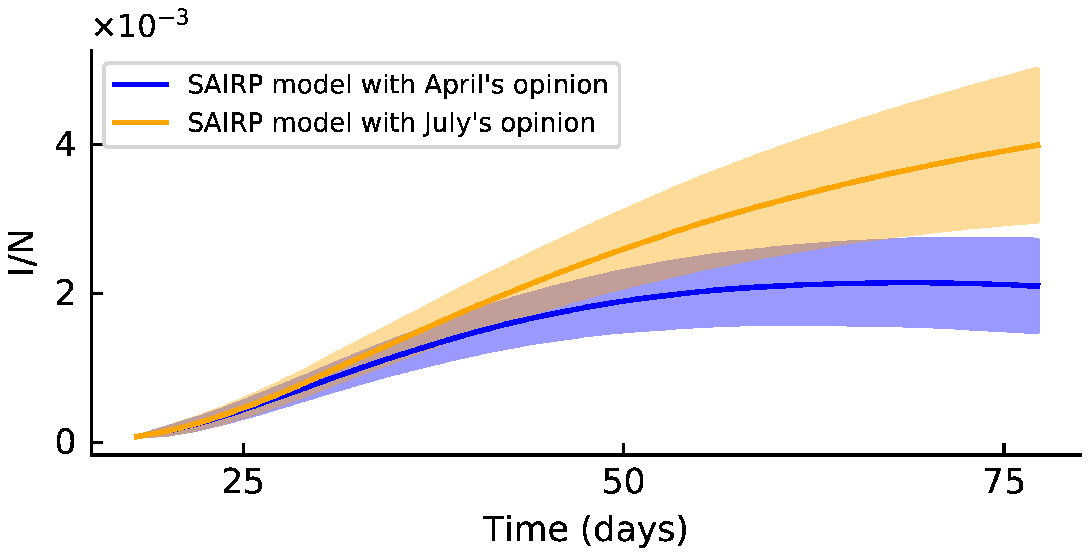}}
\caption{\textbf{Evolution of the number of infected individuals 
(normalized by the total population) with time.} (a) Red crosses correspond 
to the experimental recordings while the blue line is the fit 
of the SAIRP model with opinion. The bluish shadow marks 
the uncertainty of the model.  (b) Blue line 
is the fit of the SAIRP model coupled with the opinion distribution, 
corresponding to April 2020, and the yellow line is the evolution 
of the model coupled with the state of social opinion as in July 2020.}
\label{fig3}
\end{figure}

The results of the SAIRP model with the opinion distributions included 
are presented in Fig.~\ref{fig:opinion_expersVSApril}. The red crosses 
mark the experimental observations until May 17 and the blue line 
is the fit to the SAIRP model with the opinion distribution. The model simulation 
was repeated $12000$ times in order to gain statistical significance,
i.e., the evolution of the number of infected individuals 
shown in Fig.~\ref{fig3} is consistent and does not depend on a limited number 
of realizations, but is rather generic as the average over a significantly 
large number of simulations. The parameters used for these simulations 
are in Table~\ref{table:opinionparameters}, in the Methods section.

In Fig.~\ref{fig:opinion_AprilVSJuly}, the results of the SAIRP 
model, coupled with the opinion distributions, are shown for the 
two situations considered. The blue line corresponds to the situation 
in April 2020. The yellow line shows a possible line of evolution 
of the pandemic in case the distribution of opinion is such as in 
July 2020 (the rest of the parameters were kept as in the blue curve). 
Note that the yellow line shows a much worse scenario and it is 
a direct conclusion of a change in the distribution of opinions.


\subsection*{Optimal control}
\label{sec:oc}

We obtain optimal control strategies that respect 
the following important constraints. (i) One needs to ensure that the number 
of hospitalized individuals with COVID-19 is such that the health system 
can respond to the other diseases in the population, in order that the mortality 
associated with other causes does not increase.  
(ii) It is important that the number of active infected individuals 
is always below a critical level. (iii) In order to keep the country ``working'', 
there is always a percentage of the population that is susceptible to get infected.
For instance, it is very important to keep schools open, in particular for children under 
10/12 years old; there are always people that do not follow the rules imposed by the government; etc.
Roughly speaking, our goal is to maximize the number of people that go back to ``normal life'' 
and minimize the number of active infected (and, consequently, the number of hospitalized and 
in ICUs), ensuring that the health system is never overloaded.


\subsubsection*{Hospitals and intensive care units occupancy beds by COVID-19}
\label{sec:hosp:uci}

For the hospitalized individuals, the official data for the fraction 
of hospitalized individuals due to COVID-19, represented by $H$, 
with respect to the active infected individuals $I$ is plotted in 
Supplementary Fig.~2 (a), $H/I$. We observe that after a 
first period, where all the active confirmed cases were hospitalized, 
the so-called \emph{containment phase}, the percentage of active infected 
individuals that needs hospital treatment is always below 15\%. Moreover, 
after the end of the emergency states (red dot in Supplementary Fig. 2), 
the percentage of active infected individuals that needs to be treated at hospitals 
is less or equal than 5\% (the 15\% and 5\% are plotted with dotted blue 
lines in Supplementary Fig.~2). 

For the percentage of active infected individuals 
that need to be in intensive care units (ICU), 
we observe that (see Supplementary Fig. 2 (b)) 
the proportion of active infected individuals that requires 
medical assistance in ICU is always below than $6\%$ and, 
moreover, after the end of the state of emergency the percentage 
of active infected individuals in the ICU is always below $1\%$. 

 
\subsubsection*{Introduction of the control and its optimization}

One of the main challenges, facing countries struck by the pandemic, is 
the reopening of the economy while preserving the health of the population 
without collapsing the public health system. It is very important to keep the 
schools open (remember that children under 10/12 years old are not obliged 
to use a mask in Portugal) and prevent the economy to sink. 
Thus, there is a minimum number of people that 
need to be susceptible to infection. But we also need to account that the population 
do not always follow the rules imposed by governments. We have developed tools to quantify 
this effect and include it into the equations. With this idea in mind, we investigate 
the use of optimal control theory to design strategies for this phase of the disease. 
The goal now is to maximize the number of people transferred from class $P$ 
to the class $S$ (that helps keeping the economy alive) and, simultaneously,
minimize the number of active infected individuals and, consequently, 
the number of hospitalized and people needing ICU (in other words,
ensuring that the health system is never overloaded). 
We want to impose that the number of active infected cases is always below $2/3$ 
or 60\% of the maximum value observed up to now ($I_{\max}$). 
This condition warrants that the health system does not collapse.

\bigskip

The fraction of protected individuals $P$ that is transferred to susceptible $S$, 
is mathematically represented, in the $SAIRP$ model, by the parameter $m$. 
The class of active infected individuals $I$ is very sensitive 
to the change of the parameter $m$ (Supplementary Fig. 3). 

Taking into consideration the real official data of COVID-19 
in Portugal,\cite{dgs-covid} let
$I_{\max} = 2.5 \times 10^{-3}$ represent the maximum fraction of active infected 
cases observed in Portugal from March 2, 2020 until July 29, 2020. 
Note that for $m \geq 0.25$ the constraint $I(t) \leq  0.75 \times I_{\max}$ 
is not satisfied for the uncontrolled model \eqref{eq:model}. 
This means that the need of hospital beds and ICU beds can take vales 
such that the Health System can not respond, 
so we take the maximum value $I_{\max}$ as a reference point for the state constraints 
imposed on the optimal control problem, in order to ensure that in a future second epidemic 
wave the number of active infected cases remains 
below a certain percentage of this observed maximum value.

The parameter $m$ in the $SAIRP$ model, is replaced by a control function $u(\cdot)$. 
We formulate mathematically this optimal control problem and solve it (see Methods). 

\bigskip

The control function $u$ takes values between $0$ and $u_{\max}$, with $u_{\max} \leq 1$. 
When the control $u$ takes the value $0$ there is no transfer of individuals 
from $P$ to the class $S$; when $u$ takes the value $u_{\max}$, then $u_{\max} \%$ 
of individuals in the class $P$ are transferred to the class $S$ 
at a rate $w$ (see Table~\ref{table:parameters} in Methods for the meaning of parameter $w$). 

We consider a time window of 120 days. In the Supplementary Information, 
we analyze with more detail the optimal control problem subject to 
$I \leq 2/3 \times I_{\max}$ and $u_{\max} \leq 0.95$ 
(see Supplementary Figs. 4--6 and Supplementary Table~1). 

\begin{remark}
The optimal control problem under 
the state constraint $I \leq 2/3 \times I_{\max}$
is associated with a solution that implies a substantial and important difference
on the number of hospital beds occupancy and in intensive care units with respect
to the optimal control problem subject to the state constraint $I \leq 0.60 \times I_{\max}$.
The choice of the constraints  $I \leq 2/3 \times I_{\max}$ and $I \leq 0.60 \times I_{\max}$ 
comes from the mathematical numerical simulations carried out and the number of hospitals beds 
that the Portuguese Health System has available for COVID-19 assistance.
\end{remark}

\begin{figure}[!htb]
\subfloat[]{\label{fig:Inf:max06}
\includegraphics[scale=0.5]{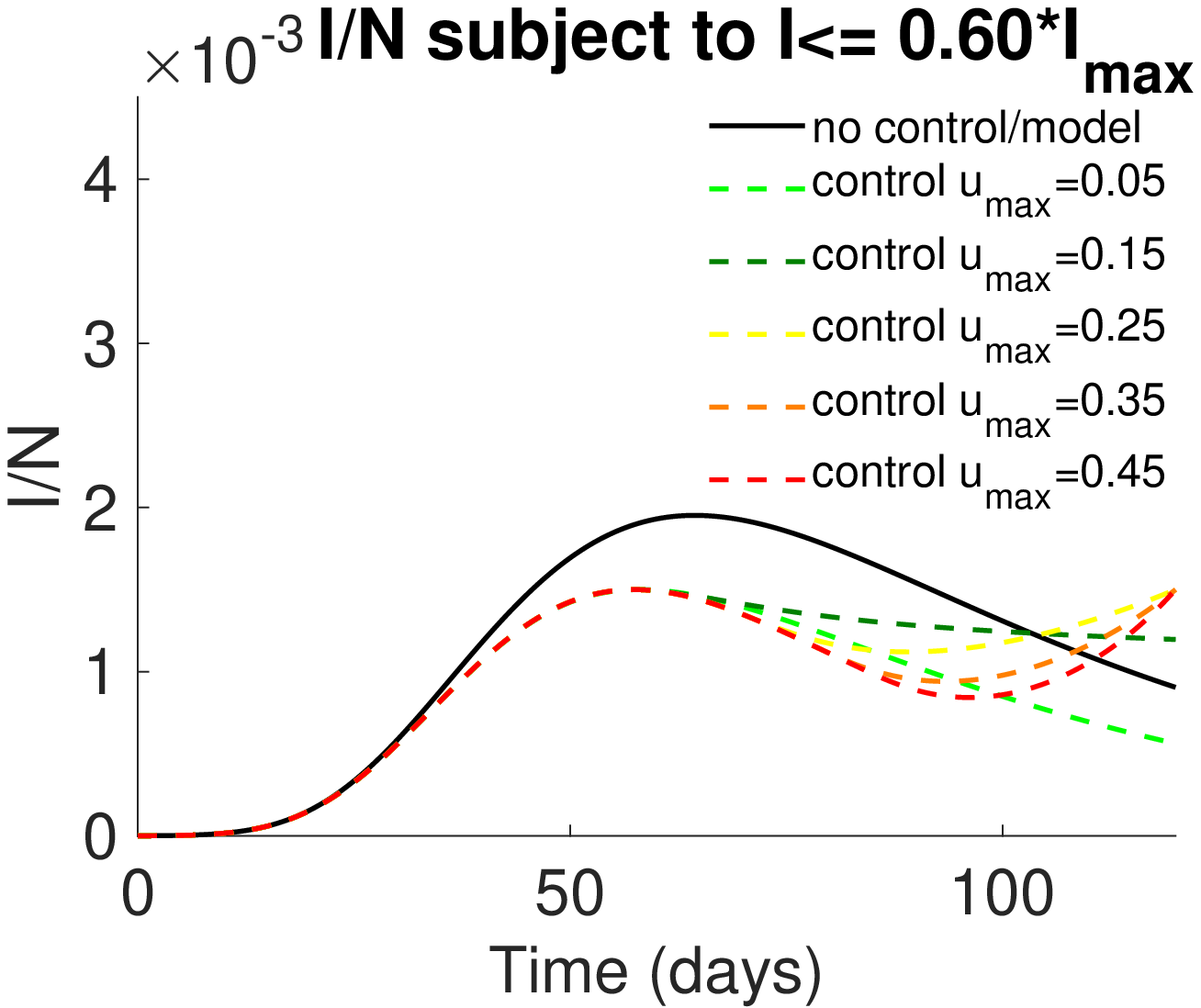}}
\subfloat[]{\label{fig:cont:max06}
\includegraphics[scale=0.5]{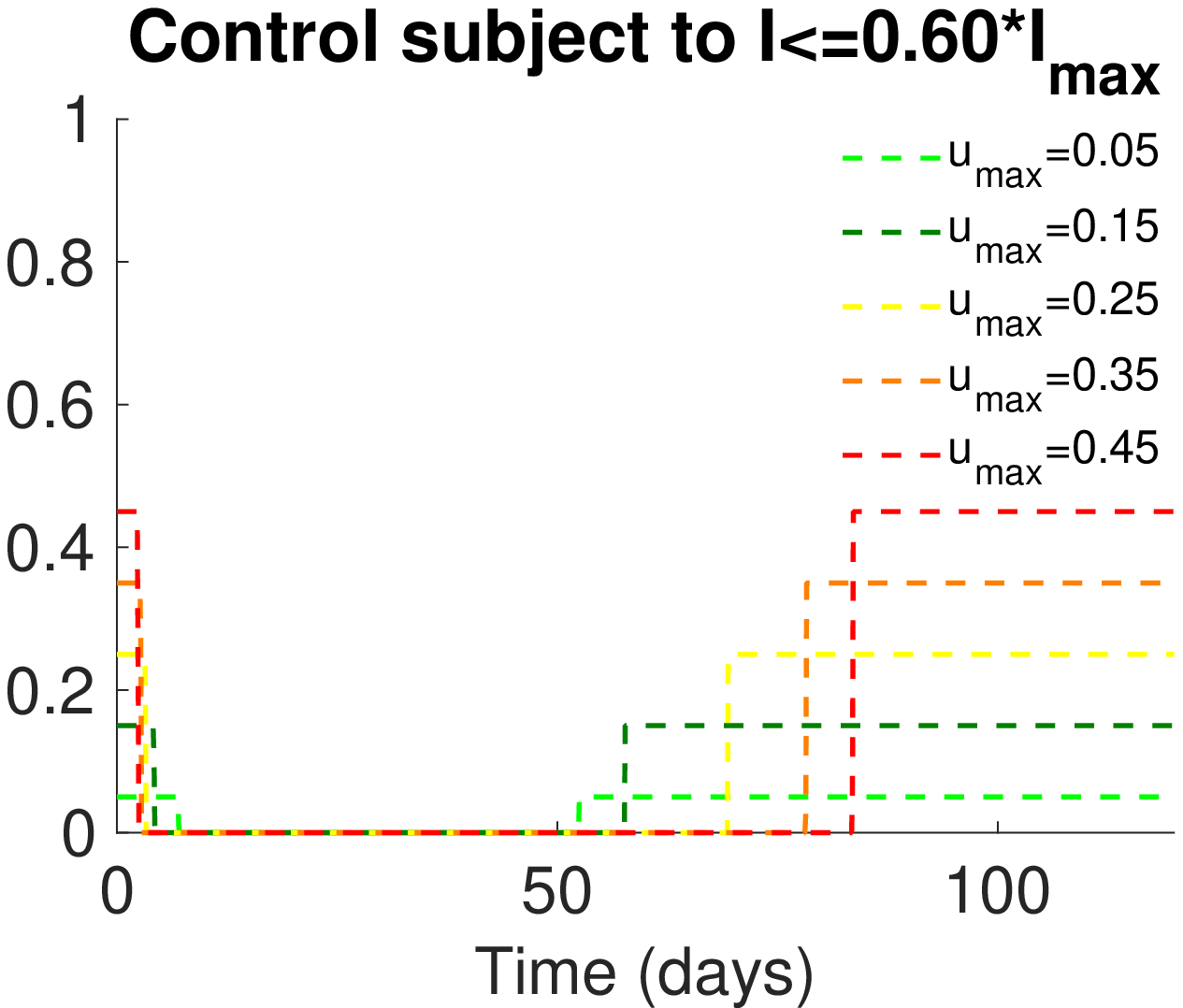}}\\
\subfloat[]{\label{fig:timePtoS:05}
\includegraphics[scale=0.5]{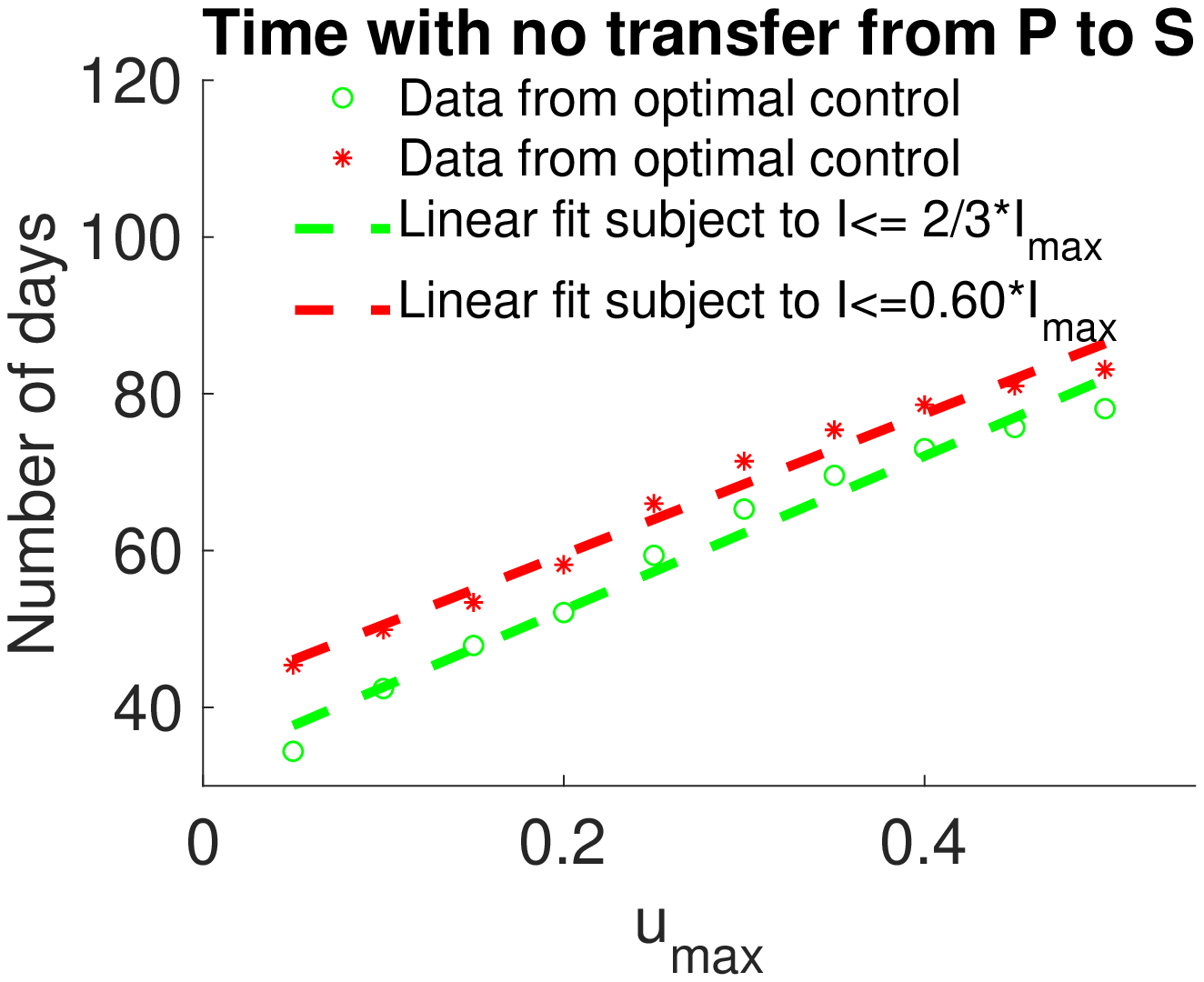}}
\subfloat[]{\label{fig:timePtoS:095}
\includegraphics[scale=0.5]{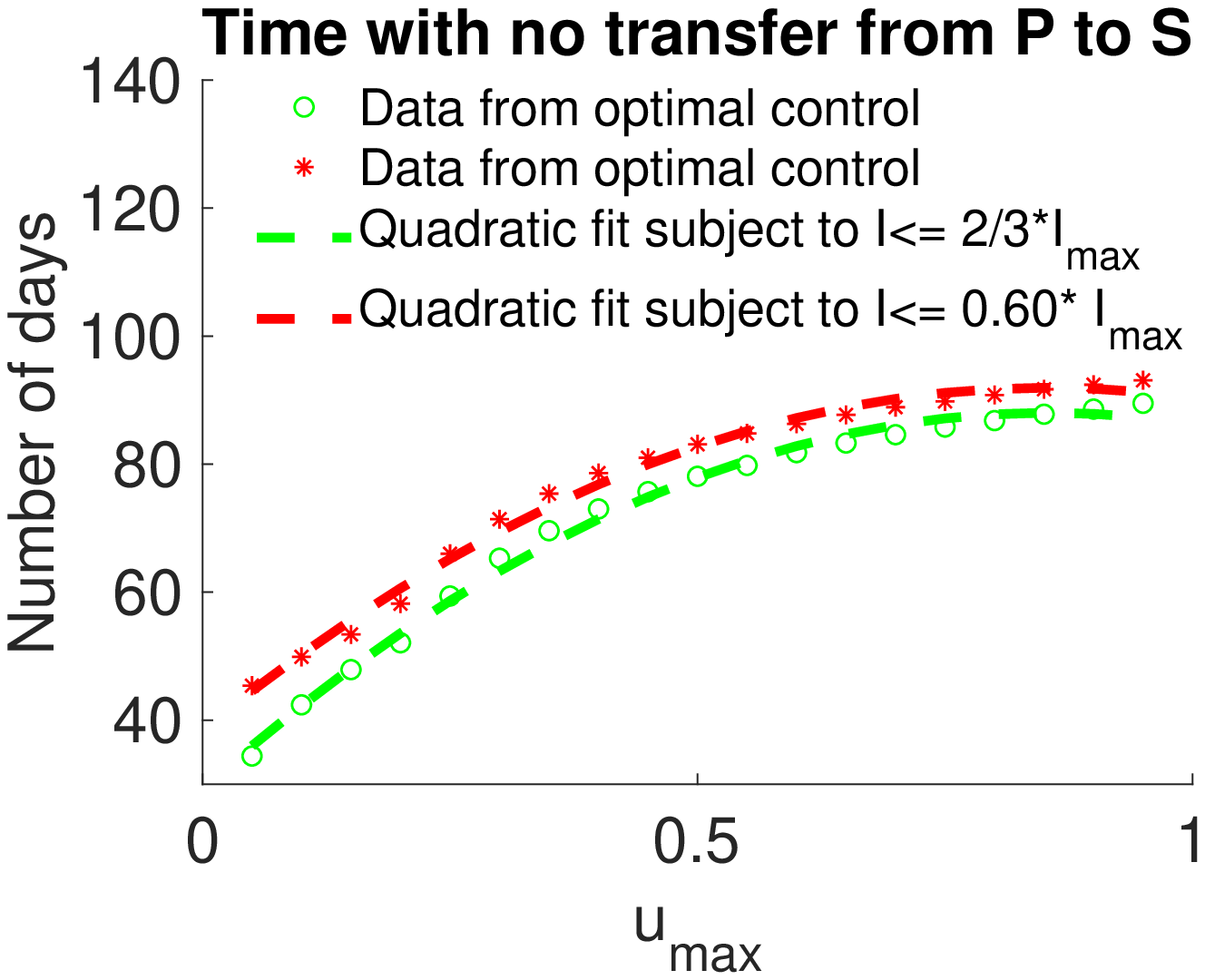}}
\caption{\textbf{Active infected individuals: comparison of the solution of the SAIRP model 
with the optimal control problem.} Linear and quadratic fit for the time where there is no 
transfer from $P$ to $S$, in terms of $u_{\max}\in[0.05;0.95]$ under the constraints 
$I \leq 0.60 \times I_{\max}$ and $I \leq 2/3 \times I_{\max}$. 
(a) Fraction of active infected individuals. (b) Control $u$ satisfying 
the constraint $I(t)\leq 0.60 \times I_{\max}$. (c) Linear fit for the 
time with no transfer from $P$ to $S$ for $0 < u_{\max} \leq 0.5$. 
(d) Quadratic fit for the time with no transfer from $P$ to $S$ for $ 0 < u_{\max} \leq 0.95$.}
\end{figure}

The controlled solution takes the maximum value $u_{\max}$ in a first period of time, 
followed by a period where there are no transfer of individuals from the class $P$ 
to the class $S$ and, at the final period of time, it takes the maximum value 
again (Fig.~\ref{fig:Inf:max06}-\ref{fig:cont:max06}). The case $u_{\max} > 0.5$ 
corresponds to a large number of days where there is no transfer of individuals 
from the class $P$ to $S$ (Fig.~\ref{fig:timePtoS:05}--\ref{fig:timePtoS:095} 
and Supplementary Fig.~5).

The time with no transfer from $P$ to $S$ corresponds to a window of time where 
strict rules are imposed to the population, that can include home confinement, for example. 
This interval of time increases when the maximum value of the control $u_{\max}$ increases 
(see Supplementary Figs.~7 and 8). We are able to compute the absolute number of individuals 
that are released to the class $S$ in terms of $u_{\max}$, which is a strictly 
increasing function of time (see Supplementary Fig.~9). 

Without loss of generality, in what follows we consider $0 < u_{\max} \leq 0.25$, 
and analyze the hospital bed occupancy and ICU beds, due to COVID-19, associated 
to the optimal solutions that satisfy the constraint $I \leq 0.60 \times I_{\max}$ 
(see Fig.~5). For the bed occupancy due to COVID-19, we give information about the number 
of total beds needed in the cases where the percentage of active infected individuals 
that needs hospital care was between 5\% and 15\% (see Fig.~5 (a)).  
This number is relatively small for the Portuguese capacities 
and will allow the medical assistance for non COVID-19 diseases.
\begin{figure}[!htb]
\centering
\subfloat[]{\label{fig:Bed:ocu:fill}
\includegraphics[scale=0.5]{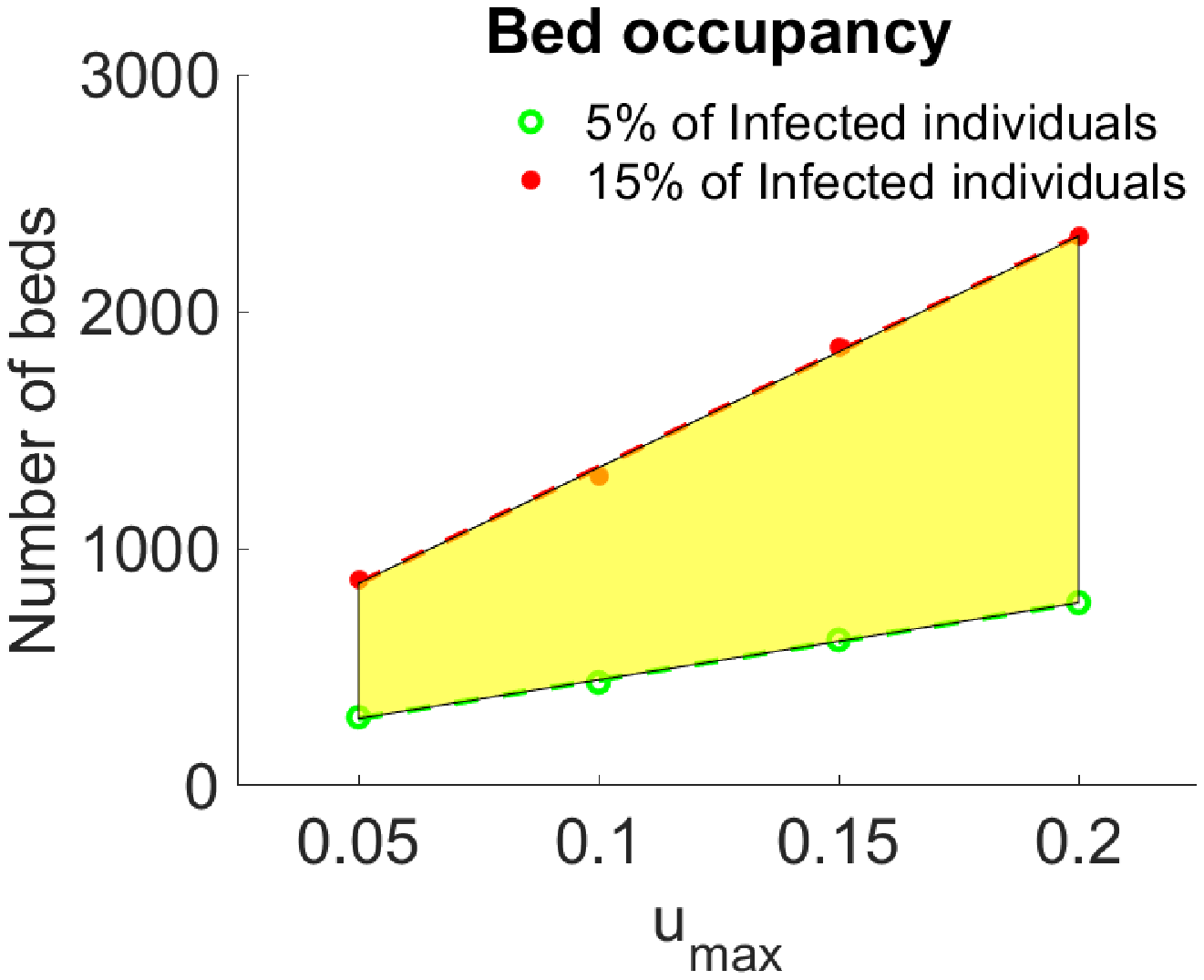}}\\
\subfloat[]{\label{fig:Bed:ocu}
\includegraphics[scale=0.4]{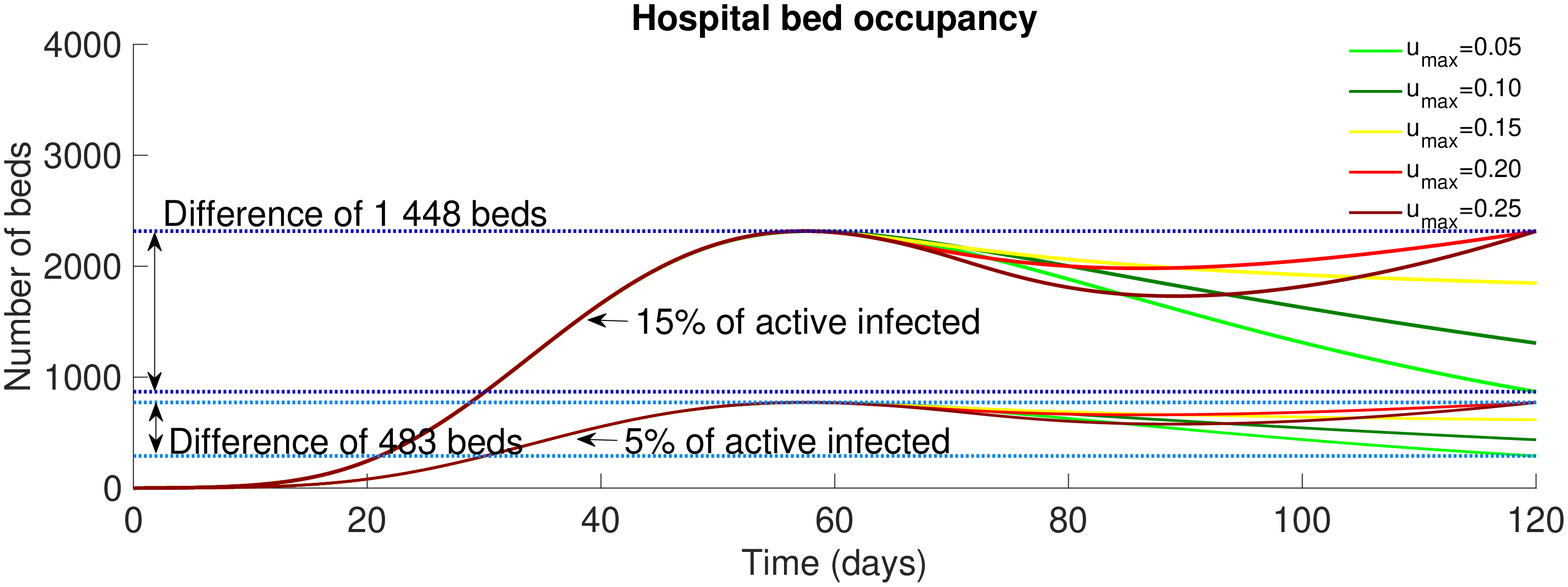}}	\\
\subfloat[]{\label{fig:ICU:Bed:ocu}
\includegraphics[scale=0.4]{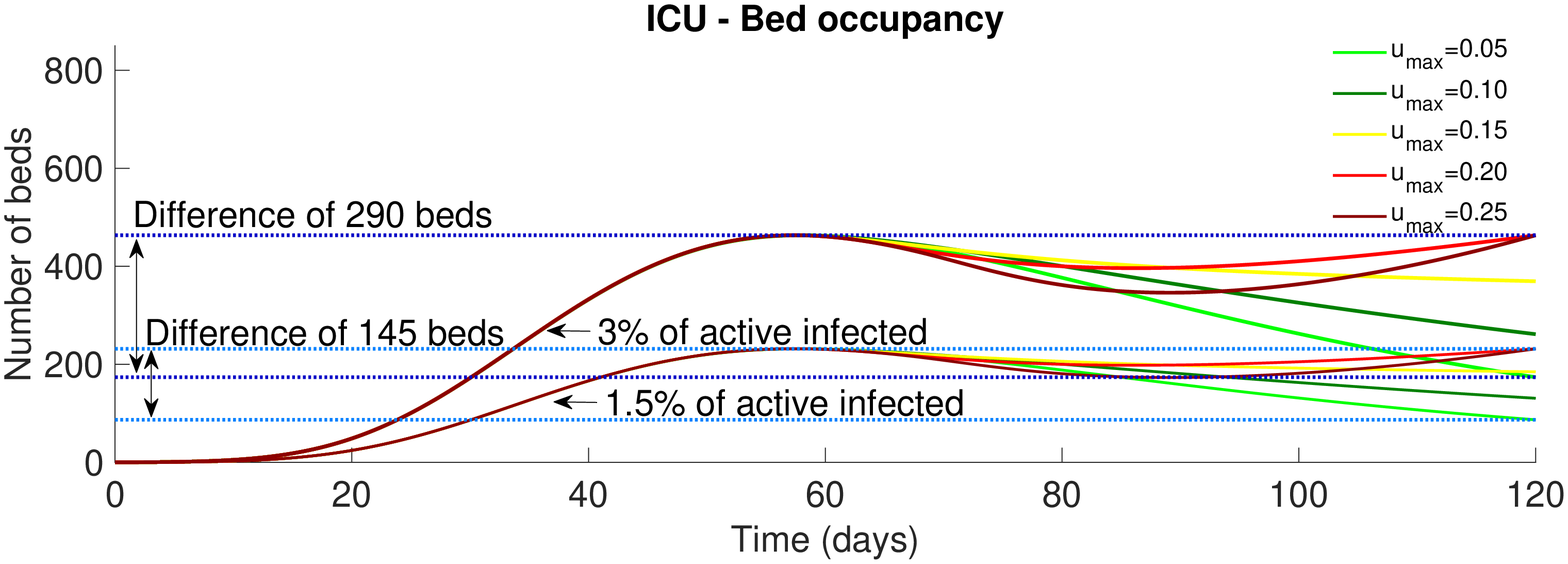}}	\\
\caption{\textbf{Number of hospital beds occupation for the optimal control solutions.}  
(a) Number of hospital beds for $u_{\max}\in\{0.05, 0.10, 0.15, 0.20\}$ 
subject to $I(t)\leq 0.60 \times I_{\max}$ varying between 
5\% and 15\% of the number of infected individuals. 
(b) Number of hospital beds for $u_{\max}\in\{0.05, 0.10, 0.15, 0.20, 0.25\}$ 
under the state constraint $I(t)\leq 0.60 \times I_{\max}$, 
representing between 5\% and 15\% of the number of active infected individuals. 
(c) ICU hospital bed occupancy for $u_{\max}\in\{0.05, 0.10, 0.15, 0.20, 0.25\}$ 
under the state constraint $I(t)\leq 0.60 \times I_{\max}$. 
The ICU beds occupation represents between 1.5\% and 3\% of the number of active infected individuals.}
\label{fig:hosp:bed:control}
\end{figure}
Considering a range of values for maximum value of the percentage of \emph{protected} 
individuals that is transferred to the \emph{susceptible} between $0.05$ and $0.25$, 
that is $u_{max} \in \{  0.05, 0.10, 0.15, 0.20, 0.25 \}$, the number of hospital beds 
needed to treat COVID-19 patients has a variation of 1448 beds, in the case when 15\% 
of active infected individuals need medical assistance (see Fig.~5 (b)). 
For the ICU bed occupancy, in the case where 3\% of the active infected individuals 
require to be in ICU, the number of beds is presented in Fig.~5 (c) 
and it may differ of 290 beds, when $u_{\max}$ varies from $0.05$ to $0.25$.  


\section*{Discussion} 
\label{sec:disc}

Portugal is a country that felt naturally isolated 
during most of the quarantine, so the data 
were not disturbed by spurious influences from other countries.
Moreover, the disease was quite controlled at all times,
the distribution of the population, as well as the distribution 
of social classes, is quite homogeneous countrywide and, thus, 
mathematical models are better suited to an analysis in a country 
like Portugal. Since the society behaves quite homogeneously across 
the country, we claim the social analysis here included to be 
quite relevant. To the best of our knowledge, 
this is the first work to investigate the reality of COVID-19 
in Portugal and suggesting control measures coming from 
the mathematical theory of optimal control. 

Optimal control theory is a branch of mathematics that offers 
a tool to tackle the problem of finding optimal strategies 
to stop the transmission of SARS-CoV-2. It is a powerful 
tool to design control strategies and act optimally 
on a given system. Based on reliable mathematical models 
for transmission mechanism of COVID-19, mathematical optimal
control can thus help and assist the Public Health Authorities to understand, 
anticipate and mitigate the spread of the virus, and evaluate the potential 
effectiveness of specific prevention strategies. 

A compartmental deterministic model describing the course 
of the epidemic, using data from Italy during the first 46 days 
(from February 20 through April 5, 2020), concluded that 
``restrictive social-distancing measures will need to be combined 
with widespread testing and contact tracing to end 
the ongoing COVID-19 pandemic'' \cite{Giordano:modcovid:NatMed}.
This has been implemented in Portugal. A stochastic microsimulation 
agent-based model of the SARS-CoV-2 epidemic for France concluded that 
``lockdown is effective in containing the viral spread, once lifted, 
regardless of duration, but it would be unlikely to prevent a rebound.'' 
The model calibrated well, based on a visually good fit between observed 
and model-predicted daily ICU admissions, ICU-bed occupancy, daily mortality 
and cumulative mortality \cite{Hoertel:modFrance:Natmed}. Our model goes further; 
it does the fit of active infected individuals and, based on that, 
estimates the number of hospitalized individuals 
with COVID-19 and the ones that are in ICU.
A projection of the SARS-CoV-2 transmission dynamics through a postpandemic period,
has been carried out with the help of a SEIR model with two strains \cite{Kissler860}.
For that, time-series data from USA has been used to calibrate the 
SARS-CoV-2 transmission model \cite{Kissler860}. They concluded that a prolonged 
or intermittent social distancing may be necessary into 2022, with additional interventions, 
including expanded critical care capacity and an effective therapeutic,
for the acquisition of herd immunity to be possible \cite{Kissler860}.  
Instead of recommending the expansion of care capacity, here we propose measures 
that maintain the number of active cases in a low level. Teslya et al. \cite{Teslya:PLOSMed2020},
suggest that information dissemination about COVID-19, which causes individual 
adoption of hand-washing, mask-wearing, and social distancing, can be an effective 
strategy to mitigate and delay the epidemic, stressing the importance of disease awareness 
in controlling the ongoing epidemic and recommending, in addition to policies 
on social distancing, that governments and public health institutions mobilize 
people to adopt self-imposed measures with proven efficacy in order 
to successfully tackle COVID-19. This was the case
in Portugal. The Portuguese experience, which prevented the rupture of the national health system, 
shows that health literacy should be a central objective at reach. Before political 
power closed schools and other institutions, the community anticipated 
and it took preventive measures. In our study, more than that, we use optimal control and network 
theories with social opinion to enrich such efforts. 
Although many other mathematical models have been already
proposed for COVID-19, the model we introduce here allows to represent, with a good fit, 
the fraction of active infected individuals in Portugal, for more than 150 days, 
and provides an interesting balance between much more complex models, 
with several more compartments, and the too much simplistic SIR/SEIR models.
Furthermore, in this work we do not simply study the sensitivity of the model 
to the change of the fraction of individuals that is in the protected class
and goes back to the susceptible, which can be done by changing some parameter values, 
but we propose optimal control solutions. 

In many countries, Portugal included, the so-called non-pharmaceutical 
interventions (NPIs) were taken since the first confirmed case. Therefore, 
our mathematical model considers a class of individuals that practice, 
in an effective way, the NPIs measures and, therefore, is 
\emph{protected} from the virus. Based on recent studies 
\cite{protective:Lancet,Haug:NatHB2020}, 
we assume that the individuals that follow NPIs measures are protected from infection 
of SARS-CoV-2. It is important to keep people in the class of protected/prevented 
due to the existing risk of transmission of the infection by asymptomatic 
infected individuals \cite{Moghadas}. 	
	
We propose a $SAIRP$ mathematical model, that represents 
the transmission dynamics of SARS-CoV-2 in a homogeneously mixing 
constant population. The $SAIRP$ model fits the confirmed active 
infected individuals in Portugal, from the first confirmed case, 
on March 2, 2020, until July 29, 2020, using real data from 
Portuguese National Authorities \cite{dgs-covid}. 
The new model considers a class of individuals that we call \emph{protected/prevented}, 
representing the fraction of individuals that is under effective protective measures, 
preventing the spread of SARS-CoV-2. In a first phase, from March 14 until May 02, 2020, 
this class represented all the individuals that were in confinement, due to closed schools, 
layoff, etc. After the three states of emergency implemented in Portugal, 
the confinement measures started to be raised but, simultaneously, 
other prevention measures were recommended by the Government, such as the use 
of mask, that became mandatory in closed spaces. All the individuals 
that practice, in a effective way, all NPIs, 
are considered to belong to class $P$. The social opinion network
implemented shows how the Portuguese population has followed the 
health authorities policies and recommendations: social distance, use of mask, 
avoid of celebrations, etc. In practice, this can be related to the partial
maintenance of the population in the class $P$ of the $SAIRP$ model. However, 
there is always a significant percentage of the population that does not follow, 
in an effective way, the official recommendations. Moreover, 
there are groups in the population that are crucial to a ``normal life'' 
and cannot avoid close physical and unprotected contacts, such as 
children in kindergartens and primary schools. With this background, 
we formulate an optimal control problem, where the control represents 
the percentage of protected/prevented individuals that are transferred 
to the susceptible class, that is, is not under protective measures.
The goal to consider such optimal control problem is to find the optimal strategy 
to transfer individuals from \emph{protected/prevented} class to the class of \emph{susceptible}, 
with minimal active infected individuals and always below a specific threshold 
that maintains the number of hospitalized individuals due to COVID-19 and 
hospitalized in intensive care units, below the level that the National Health Service 
is able to answer while keeping the other ``usual'' medical services working normally. 
This is also connected with the political and social interest of keeping the economy 
open and ``active''. We provide the mathematical optimal control solutions 
for different scenarios on the fraction of protected individuals that is transferred 
to the susceptible class and also for different threshold levels. 

We conclude with some words explaining why we believe optimal control
has an important role in helping to prevent COVID-19 dissemination,
and also pointing out some possible future research directions.
In general, the response to chronic health problems has been impaired, 
both because the resources were largely allocated to COVID-19 
or because the population was afraid to go to the hospitals 
and many surgeries and consultations remain to be made. 
Many institutions have organized what has been called in Portugal
``home hospitalization'', which served to mitigate many
problems that would remain unanswered. Hospital teams,
multidisciplinary teams, systematically moved to the homes of
patients and sought care in their environment, avoiding nosocomial 
infections and also the occupation of beds. This experience 
was evaluated as very positive by the Portuguese population.
Most probably, this coronavirus will remain in the communities 
for many years, so the changes we see in health services 
and in people's habits have to go on over time. Actions 
as simple as hand washing, space hygiene, social distance 
and use of masks in closed spaces, should be incorporated 
into education for health. The containment measures, 
which should be necessary when outbreaks arise, 
must be rigorously studied and worked with families. 
Confinement cannot mean social isolation
and should be worked out according to each family reality.
The latest data shows that European countries are already at the limit 
in terms of reinforcements to NHS budgets. Changing many hospital practices, 
such as cleanliness and hygiene, food services, relationship between
emergencies and hospitalization, support for clinical training 
of health professionals, etc., can help to rationalize resources 
and prevent infections to other users, especially in autumn and winter, 
where different forms of flu and pneumonia burden institutions.
At this moment we do not include such ``social'' corrections 
in the optimal control part, but it would be interesting 
to consider them in future work.


\section*{Methods}
\label{sec:methods}


\subsection*{Mathematical epidemiological model}

The SAIRP model \eqref{eq:model} subdivides human population 
into five mutually-exclusive compartments (see Table~\ref{table:compart:model} 
and Supplementary Fig.~1), representing the dynamical evolution of the population 
in each compartment over a fixed interval of time.  
\begin{table}[!htb]
\caption{Description of the population model compartments.}
\label{table:compart:model}
\centering
\begin{tabular}[center]{ l l } \hline
Population compartment  &  Description \\ \hline 
$S$ & susceptible \\	
$A$ & asymptomatic \\
$I$ & confirmed/active infected \\
$R$ & recovered/removed (includes deaths by COVID-19) \\
$P$ & protected/prevented  \\ \hline	
\end{tabular}
\end{table}
The susceptible individuals become infected by SARS-CoV-2 by contact 
with infected asymptomatic $A$ and active infected individuals $I$. 
The rate of infection is given by $\beta \left( \theta A(t) + I(t) \right)$, 
where $\beta$ is the infection transmission rate of active infected individuals 
$I$ and $\theta$ represents a modification parameter for the infectiousness 
of the asymptomatic infected individuals ($A$). A fraction $p$, with $0 < p < 1$, 
is protected from infection by SARS-CoV-2, due to an effective implementation 
of non-pharmaceutical interventions (NPIs) and is transferred to the class $P$, 
at a rate $\phi$. However, individuals in the class $P$ are not immune to 
infection and a fraction $m$ can become susceptible again at a rate $w$. 
For the sake of simplification, we denote $\omega = w m$.    
A fraction $q$ of asymptomatic infected individuals $A$ develop symptoms 
and are detected, at a rate $v$, being transferred to the class $I$. 
We use the notation $\nu = v q$. Active infected individuals $I$ exit this class 
either by recovery from the disease or by COVID-19 induced death, 
being transferred to the class of removed/recovery $R$, 
at a rate $\delta$ (see Table~\ref{table:parameters}).   
\begin{table}[!htb]
\caption{Description of the parameters of model \eqref{eq:model}. }
\label{table:parameters}
\centering
\begin{tabular}[center]{ l l } \hline
Parameter/  &  Description  \\ \hline 
$\beta$ & Infection transmission rate \\ 
$\theta$ & Modification parameter \\
$p$ & Fraction of susceptible $S$ transferred to protected class $P$\\
$\phi$ & Transition rate of susceptible $S$ to protected class $P$\\
$\omega = w m$ & \\
$w$ & Transition rate of protected $P$ to susceptible $S$ \\
$m$ & Fraction of protected $P$ transferred to susceptible $S$\\
$\nu = v q$ & \\
$v$ & Transition rate of asymptomatic $A$ to active/confirmed infected $I$\\
$q$ & Fraction of asymptomatic $A$ infected individuals \\
$\delta$ & Transition rate from active/confirmed infected $I$ to removed/recovered $R$\\ \hline 
\end{tabular}
\end{table}
The previous assumptions are described by the following system of five ordinary differential equations: 
\begin{equation}
\label{eq:model}
\begin{cases}
\dot{S}(t) =  - \beta (1-p) \left( \theta A(t) + I(t) \right) S(t) - \phi p S(t) + \omega P(t)  ,\\[0.2 cm]
\dot{A}(t) = \beta (1-p) \left( \theta A(t) + I(t)  \right) S(t) - \nu  A(t) , \\[0.2 cm]
\dot{I}(t) = \nu A(t) - \delta  I(t) ,\\[0.2 cm]
\dot{R}(t) = \delta I(t), \\[0.2 cm]
\dot{P}(t) = \phi p S(t) - \omega P(t) . 
\end{cases}
\end{equation}

\begin{remark}	
The testing rate in Portugal, as in many other countries, 
has been increasing since the beginning of the pandemic. However, in our model 
we do not consider the impact of the testing rate on the detection of infected cases. 
This is due to the fact that in Portugal only suspected individuals that had 
a close contact, without mask protection, or individuals with COVID-19 symptoms, 
are tested.
\end{remark}

\begin{remark}	
In our model \eqref{eq:model}, 
individuals from compartment $A$ move to compartment $I$. 
Given testing frequencies and reliability, we adjust the infection rate 
and consider a proportion of detection. The parameter $q$ is used to obtain 
the proportion of $A$ moving to $I$. In a general framework, a fraction $(1-q)$ 
of asymptomatic individuals $A$ should be transferred to the compartment $R$. 
However, in this work we are based on the official data provided by 
The Portuguese Health Authorities and our aim is to propose a mathematical 
model that fits well the reality described by the daily reports data, 
more specifically the curve of the active infected individuals by COVID-19 in Portugal and, 
sub-sequentially, the fraction of active individuals that are hospitalized and in intensive care units. 
Using official data, only the individuals that were confirmed to be infected by testing 
(the ones that are represented by the class $I$) may be transferred to the class $R$. 
Therefore, since the asymptomatic are not counted in the official data, 
it is not possible (in this model) to count them as recovered after a certain number of days.
\end{remark}

Let us define the total population $N$ by 
$N(t) = S(t) + A(t) + I(t) + R(t) + P(t)$. Taking the derivative
of $N(t)$, it follows from \eqref{eq:model} that $\dot{N}(t) = 0$, 
that is, $N$ is constant over time. Without loss of generality,
we normalize the system so that $N = 1$. All parameters of the model 
are non-negative and, given non-negative 
initial conditions $(S_0, A_0, I_0, R_0, P_0)  
= \left( S(0), A(0), I(0), R(0), P(0)\right)$, 
the solutions of system \eqref{eq:model} are non-negative 
and satisfy $S(t)+A(t)+I(t)+R(t)+P(t) = 1$ for all time $t \in [0, t_f]$.
With this conservation law, the model \eqref{eq:model} can be simplified 
to 4 equations, the cumulative number of removed/recovered individuals 
$R(t)$ being given, for each $t \geq 0$, by
\begin{equation}
\label{eq:R}
R(t) = R(0) + \delta \int_{0}^{t} I(s) \, ds \, . 
\end{equation}
Therefore, we consider the following $SAIP$ simplified model
for the optimal control problem formulation:
\begin{equation}
\label{eq:model:2}
\begin{cases}
\dot{S}(t) =  - \beta (1-p) \left( \theta A(t) 
+ I(t) \right) S(t) - \phi p S(t) + \omega P(t),\\[0.2 cm]
\dot{A}(t) = \beta (1-p) \left( \theta A(t) 
+ I(t)  \right) S(t) - \nu  A(t) , \\[0.2 cm]
\dot{I}(t) = \nu A(t) - \delta  I(t) ,\\[0.2 cm]
\dot{P}(t) = \phi p S(t) - \omega P(t) . 
\end{cases}
\end{equation}

\begin{remark}
In our model we are taking into account the infectiousness 
of fully asymptomatic patients $A$. In concrete, 
the transmission incidence is given 
by the term $\beta (1-p) (\theta A(t) + I(t))$.
\end{remark}

The disease free equilibrium $\Sigma_0$ 
of model \eqref{eq:model:2} is given by
\begin{equation}
\label{DFE}
\Sigma_0 = \left\{ S=\frac{\omega}{\phi\,p+\omega}, \, 
A=0, \, I=0, \, P=\frac {\phi\,p}{\phi\,p+\omega} \right\}
\end{equation}
with $S + P = 1$. Following the approach of Driessche and Watmough, \cite{Driessche} 
the basic reproduction number $R_0$ is given by the spectral radius 
of $FV^{-1}$, where the matrices $F$, $V$ and $FV^{-1}$ are given by 
\begin{equation*}
F = 
\begin{pmatrix}
-S\beta(p-1)\theta & -\beta(p-1)S \\[2ex]
0 & 0 
\end{pmatrix}
,\qquad
V = 
\begin{pmatrix}
\nu & 0 \\
-\nu & \delta 
\end{pmatrix} \, ,
\end{equation*}
\begin{equation*}
FV^{-1} = 
\begin{pmatrix}
-{\frac {\beta(p-1)\theta\omega}{(\phi p+\omega)\nu}}
-{\frac {\beta(p-1) \omega}{(\phi p+\omega) \delta}}\quad & 
-{\frac {\beta(p-1) \omega}{( \phi p+\omega) \delta}} \\[2ex]
0 & 0 
\end{pmatrix},
\end{equation*}
that is,
\begin{equation}
R_0 = \frac{\beta(1-p)\, \omega \, (\theta\delta+\nu)}{(\phi p+\omega)\, \nu \,\delta} \, .
\end{equation}


\subsection*{Parameter values and estimation from Portuguese COVID-19 data}
\label{sec:fit:justication}

We consider official data, where daily reports are available with the information about total 
(cumulative) confirmed infected cases, total recovered, 
and total deaths by COVID-19 in Portugal, and also information about the number 
of hospitalized individuals and in intensive care due to COVID-19 disease.\cite{dgs-covid} 
\begin{table}[!htb]
\caption{\textbf{Initial conditions and parameter values for Portugal 
from March 2, 2020 to June 19, 2020.} Contrast with Fig.~\ref{fig:Inf}. 
The parameters $\beta_1$, $\beta_2$ and $\beta_3$, 
and $m_1$, $m_2$ and $m_3$ were estimated using 
the \textsf{Matlab} function \texttt{lsqcurvefit} for $t \in [0, 77]$ and 
$t \in [100, 150]$ days, respectively. From  March 2 to May 17, 2020 (77 days): 
$t \in [0, 77]$ -- $\beta_1 = 1.492$, $m_1 = 0.059$ and $p_1 = 0.675$. 
From May 17 to June 9, 2020 (23 days): $t \in [77, 100]$ -- 
$\beta_2 = 0.25$, $m_2 = 0.058$ and $p_2 = 0.4$.
From June 9 to July 29, 2020 (50 days): $t \in [100, 150]$ 
-- $\beta_3 = 1.91$, $m_3 = 0.043 $ and $p_3 = 0.4$.}	  
\label{table:param:values:PT}
\centering
\begin{tabular}[center]{ l l l }  \hline
Parameter/Initial condition  & Value & Reference  \\  \hline 
$\beta_1$ & $1.492$ & Estimated \\ 
$\beta_2$ & $0.25$ & Estimated \\ 
$\beta_3$ & $1.91$ & Estimated \\ 
$\theta$ &  $1$ & \cite{cdc:param:est} \\
$p_1$ & $0.675$ & \cite{dgs-covid,legislacao:covid19} \\  
$p_2$ & $0.55$  &   \\
$p_3$ & $0.40$  &  \\
$\phi$ &  $1/12 \,  day^{-1}$ & \cite{legislacao:covid19,AnaPaiao:Ecology2020}\\
$\omega = w m$ &  & \\
$w$ & $1/45  \, day^{-1}$ & \cite{legislacao:covid19}\\
$m_1$ & $0.059$ &  Estimated \\
$m_2$ & $0.058$ &  Estimated\\
$m_3$ & $0.043$ &  Estimated \\
$\nu = v q$ &  & \\
$v$ & $1   \, day^{-1}$ & \\
$q$ & $0.15$ & \cite{Li:Science:2020,Mizumoto:2019,Park:Epidemics:2020} \\
$\delta$ & $1/30  \, day^{-1}$  & \cite{lancet:timerecover:2020} \\ \hline	
$N = S_0 + A_0 + I_0 + R_0 + P_0$  & $10 295 909$ & \cite{INE} \\
$S_0$ & $10295894/N$  & \cite{dgs-covid} \\
$I_0$  &  $2/N$ &  \cite{dgs-covid}  \\  
$A_0$ &  $(2/0.15)/N$ & \cite{dgs-covid} \\	
$R_0$ & $0$  & \cite{dgs-covid}\\
$P_0$ & $0$  & \cite{dgs-covid} \\ \hline 
\end{tabular}
\end{table}
We assume $\theta = 1$ for the current (up to the date) 
best estimate for the infectiousness of asymptomatic 
individuals relative to symptomatic individuals.\cite{cdc:param:est} 
For the fraction of asymptomatic $A$ infected individuals, we consider 
$q = 0.15$.\cite{Li:Science:2020,Mizumoto:2019,Park:Epidemics:2020} 
The parameter $w$ takes the value $w=1/45  \, day^{-1}$, 
corresponding to the 3 emergency states (duration 45 days) \cite{legislacao:covid19}.
The value of the parameter $\delta$, representing the recovery time of confirmed active 
infected individuals $I(t)$ (with negative test)/removed (by death), is assumed to be 
$\delta = 1/30 \, \text{days}^{-1}$, considering that here might 
be a delay on the publication of real data.\cite{lancet:timerecover:2020}
The parameters $\beta_1$, $\beta_3$, $m_1$ and $m_3$ were estimated using the \textsf{Matlab} 
function \texttt{lsqcurvefit} for $t \in [100, 150]$ days, respectively. 
From March 2 to May 17, 2020 (77 days): $t \in [0, 77]$ -- $\beta_1 = 1.492$, 
$m_1 = 0.059$, and $p_1 = 0.675$. From May 17 to June 9, 2020 (23 days): 
$t \in [77, 100]$ -- $\beta_2 = 0.25$, $m_2 = 0.058$ and $p_2 = 0.4$. 
From June 9 to July 29, 2020 (50 days): $t \in [100, 150]$ --
$\beta_3 = 1.91$, $m_3 = 0.043$, and $p_3 = 0.4$. 
The fraction $0 < p_1 < 1$, for $t \in [0, 77]$, 
is assumed to take the value $p_1=0,675$, 
representing the population affected by the confinement of policies 
\cite{dgs-covid,legislacao:covid19}.  
For $t \in [77, 100]$, we assume a decrease 
of the fraction of \emph{protected} individuals to $p_2=0.55$. For $t \in [100, 150]$, 
we assume $p_3 = 0.44$, based on a gradual transfer of individuals 
from the class $P$ to the class $S$. The transfer of individuals from $S$ to $P$ 
started on March 14, 2020 \cite{legislacao:covid19,AnaPaiao:Ecology2020}, 
thus we take $\phi = 1/12 \, day^{-1}$. 


\subsection*{Building the social network}
\label{sec:methods:network}

In order to generate the social network, we use data collected from the 
micro-blogging website \textsf{Twitter}. With aid of the \textsf{Python} 
package \textsf{GetOldTweets3} \cite{python}, we were able to download 
a collection of several tweets (posts of 244 characters) attending 
to participation in a given hashtag. Merging a handful of different hashtags, 
we obtained a significant sample of users who are interacting between themselves, 
either exchanging information with replies or spreading it via what is called 
a ``re-tweet''. The more hashtags we use, the more realistic 
is the reconstruction of the social network in regards to the actual situation of the Portugal 
Twitter network. In mathematical terms, we build a complex network where users 
lie in the nodes and the directed edges represent the interactions between users. 
We are mainly interested on the structure of the interactions rather than the 
topic of the information, and thus we discard everything related to the personal 
information of the users and the content of the tweets.	

Most real world networks are changing in time, either by changes on the connectivity 
pattern or either by growth and continuous addition of new nodes. This is a key 
feature of the so called scale-free complex network \cite{Albert2002}, and it is 
a feature shared by the social network Twitter \cite {Pereira2016,Abel2011,Cataldi2010}. 
Thus, the structure of the network can drastically change from one month to another, 
and so it is important to take this point into account when building the network. 
In our data, this was accomplished by a feature of the used package, which allows 
to filter the search by date. In fact, here we were also interested in comparing 
the behavior of the social network during April, when the quarantine was imposed, 
and the social network during July, when the social distancing measures relaxed. 
The connectivity distributions for both networks are shown in Supplementary Fig.~10. 
In both cases, the topology corresponds with that of a \emph{scale free network} 
but with rather different exponents, $\gamma_{April}=2.11$ and $\gamma_{July}=1.82$. 
The significantly different exponents demonstrate the different internal dynamics 
in both cases, which are reflected in the opinion distributions. 


\subsection*{Opinion model}
\label{sec:methods:opinion}

The network topology obtained was endowed with a dynamical opinion 
set of equations for each node (actual person) that, 
combined with the information coming through the network connections, 
allowed it to produce an opinion. We considered a simple opinion model 
based on the logistic equation \cite{Verhulst1845} 
but that has proved to be of use in other contexts 
\cite{Lloyd1995,Tarasova2017,Carballosa2020}. The equations describing 
each node $i$, $i=1, \ldots, N$, are \cite{Carballosa2020}:
\begin{equation}
\frac {d u_i}{d t} = f(u_i) + d \frac{1}{k_i} \sum_{j=1}^{N} L_{ij} u_{j},
\end{equation}
where $u_i$ is the opinion of node $i$ that ranges from zero to one. 
The nonlinearity $f(u_i)$ is given by the following equation:
\begin{equation}
f(u)=u \left(A (1 - u/B) + g(1-u) \right).
\end{equation}
Each of the nodes $i$ obeys the internal dynamic given by $f(u_i)$ 
while being coupled with the rest of the nodes with a strength $d/k_i$, 
where $d$ is a diffusive constant and $k_i$ is the connectivity degree 
for node $i$ (number of nodes each node is interacting with). Note that 
this is a directed non-symmetrical network where $k_i$ means that node 
$i$ is following the tweets from $k_i$ nodes and, thus, it is being 
influenced by those nodes in its final opinion. The Laplacian matrix
$L_{ij}$ is the operator for the diffusion in the discrete space, 
$i=1,\ldots,N$. We can obtain the Laplacian matrix from the connections 
established within the network as $L_{ij}=A_{ij}-\delta_{ij} k_i$, 
being $A_{ij}$ the adjacency matrix: 
\begin{equation}
A_{ij}=
\begin{cases}
1\quad \text{if} \quad i,j \text{ are connected,} \\[0.2cm]
0\quad \text{if} \quad i,j \text{ are not connected}.
\end{cases}
\end{equation}
Now, we proceeded as follows. We considered that all the accounts 
(nodes in our network) were in their stable fixed point with a $10\%$ 
of random noise. Then a subset of the nodes was forced to acquire 
a different opinion, $u_i=1$ with a $10\%$ of random noise  
and we let the system to evolve following the above dynamical equations. 
The influence of the network made some of the nodes to shift their opinion  
to values closer to 1 that, in the context of this simplified opinion model, 
means that those nodes shifted their opinion to values closer to those 
leading the shift in opinion. This process was repeated in order to gain 
statistical significance and, as a result, it provided the probability 
distribution of nodes eager to change the opinion and adhere to the new politics. 
The parameter values used were $A=0.00001$, $B=0.1$, $g=0.001$ and $d=1.0$.


\subsection*{Parameter values for the SAIRP model with opinion distributions}
\label{sec:methods:opinionparameters}

The parameters used and the initial 
conditions are summarized in Table~\ref{table:opinionparameters}.
Once the opinion distribution was included into the SAIRP model, 
the parameters were slightly adjusted to be able to continue describing 
accurately the experimental situation. In fact, moving from an
only-time-dependent-model to the network type model we consider 
in this section implies that the whole dynamic of the system is speeded up 
as now each node has the capability to trigger the epidemic wave. 
In order to compensate this effect, a rescaling of the parameters 
controlling the temporal scale in the system, namely $\delta$, $\phi$ and $w$, 
is necessary. An estimated rescaling factor of 1.85 leaves the modified parameters 
shown in Table~\ref{table:opinionparameters}. With the opinion distribution included 
into the SAIRP model, the infection transmission rate also needs to be adjusted 
in order to continue describing accurately the experimental situation.

\begin{table}[!htb]
\caption{Initial conditions and parameter values for Portugal 
from March 2, 2020 to May 17, 2020 for the SAIRP model, 
modified by the opinion distributions.}	  
\label{table:opinionparameters}
\centering
\begin{tabular}[center]{ l l }  \hline
Parameter/Initial condition  & Value   \\  \hline 
$\beta$ & $0.2$  \\ 
$\phi$ &  $1/6.486 \,  day^{-1}$  \\
$w$ & $1/24.32 \, day^{-1}$  \\
$v$ & $1   \, day^{-1}$  \\
$q$ & $0.15$  \\
$\delta$ & $1/16.216  \, day^{-1}$   \\ \hline	
$N = S_0 + A_0 + I_0 + R_0 + P_0$  & $25000$    \\
$S_0$ & $ (N-2-2/0.15)/N$    \\
$I_0$  &  $2/N$     \\  
$A_0$ &  $(2/0.15)/N$   \\	
$R_0$ & $0$   \\
$P_0$ & $0$    \\ \hline 
\end{tabular}
\end{table}


\subsection*{Optimal control problem}
\label{sec:oc:methods}

The goal is to find the optimal strategy for letting people to go out 
from class $P$ to the class $S$ and, at the same time, minimize 
the number of active infected while keeping the class of active 
infected individuals below a safe maximum value. 

The control $u(\cdot)$ represents the fraction of individuals 
in class $P$ of \emph{protected} that is transferred to the class $S$. 
The control $u$ is introduced into the $SAIP$ model in the following way: 
\begin{equation}
\label{eq:model:cont:1}
\begin{cases}
\dot{S}(t) = - \beta (1-p) \left( \theta A(t) + I(t)  \right) 
S(t) - \phi p S(t) + w u(t) P(t)  ,\\[0.2 cm]
\dot{A}(t) = \beta (1-p) \left( \theta A(t) 
+ I(t) \right) S(t) - \nu  A(t) , \\[0.2 cm]
\dot{I}(t) = \nu A(t) - \delta  I(t) ,\\[0.2 cm]
\dot{P}(t) = \phi p S(t) - w u(t) P(t) . 
\end{cases}
\end{equation}
The control must satisfy the following constraints: $0 \leq u(t) \leq u_{\max}$ 
with $u_{\max} \leq 1$. In other words, the solutions of the problem must belong 
to the following set of admissible control functions:
\begin{equation}
\label{eq:admiss:control}
\Theta = \left\{ u, \; u \in L^1 \left([0, t_f], \mathbb{R} \right) 
\, | \, 0 \leq u(t) \leq u_{\max} \, \;\; \forall \; t \in [0, t_f] \right\} \, .
\end{equation}
Mathematically, the main goal consists to minimize the cost functional
\begin{equation}
J(u) = \int_{0}^{t_f} k_1 I(t) - k_2 \, u(t) \, dt \, , 
\end{equation}
representing the fact that we want to minimize the fraction 
of infected individuals $I$ and, simultaneously, maximize the 
intensity of letting people from class $P$ go back to class $S$. 
The constants $k_i$, $i=1, 2$, represent the weights associated 
to the class $I$ and control $u$. Moreover, the solutions of the 
optimal control problem must satisfy the following state constraint: 
$I(t) \leq \zeta$, in one case with $\zeta = 0.6 \times I_{\max}$, 
the other with $\zeta = 2/3 \times I_{\max}$.  

For the numerical simulations, we considered 
$k_1 = 100$, $k_2 = 1$ and $t_f = 120$ days. We also considered 
$(\beta, \delta) = (1.464, 1/30)$, $m = 0.09$, $p = 0.675$, 
and all the other parameters from Table~\ref{table:param:values:PT}.
Numerically, we discretized the optimal control problem to a nonlinear 
programming problem, using the Applied Modeling Programming Language (AMPL) 
\cite{AMPL}. After that, the AMPL problem was linked to the optimization solver 
IPOPT \cite{IPOPT,SilvaMaurerTorres}. The discretization 
was performed with $n = 1500$ grid points using the trapezoidal rule 
as the integration method. 


\section*{Data availability}

All of the data are publicly available and were extracted 
from \url{https://covid19.min-saude.pt/relatorio-de-situacao/}. 


\section*{Code availability}

The code is available from the authors on request. 



\section*{Acknowledgments} 
\small
This research is partially supported by the Portuguese Foundation for Science and Technology (FCT) 
within ``Project Nr.~147 -- Controlo \'Otimo e Modela\c{c}\~ao Matem\'atica da Pandemia \text{COVID-19}: 
contributos para uma estrat\'egia sist\'emica de interven\c{c}\~ao em sa\'ude na comunidade'', 
in the scope of the ``RESEARCH 4 COVID-19'' call financed by FCT, and by project UIDB/04106/2020 (CIDMA). 
Silva is also supported by national funds (OE), through FCT, I.P., in the scope of the framework 
contract foreseen in the numbers 4, 5 and 6 of the article 23, of the Decree-Law 57/2016,
of August 29, changed by Law 57/2017, of July 19. 
This research is also partially supported by the ``Instituto de Salud Carlos III 
and the Ministerio de Ciencia e Innovaci\'{o}n'' of Spain, research grant COV20/00617, 
and by Xunta de Galicia, research grant 2018-PG082. APM and AC are part of the CRETUS 
Strategic Partnership (AGRUP2015/02) and JM is part of the AeMAT Strategic Partnership 
(ED431E2018/08), both supported by Xunta de Galicia. All these programs are co-funded 
by FEDER (EU). A substantial portion of the simulations were run at the 
Centro de Supercomputaci\'{o}n de Galicia (Spain) and we acknowledge their support.  


\section*{Author contributions}

C.J.S., C.C., D.F.M.T., I.A., J.J.N., A.P.M., A.C., and J.M. 
conceived the study, formulated the mathematical model, 
incorporated the social network, formulated the optimal control problem, 
conducted the analysis, and wrote the manuscript.  
C.J.S., C.C., and D.F.M.T., performed the mathematical analysis 
of the epidemiological model, formulated and solved the optimal control problem. 
A.P.M., A.C., and J.M. defined and analyzed the social network.
R.F-P., R.P.F., E.S.S, and W.A. provided the clinical contextualization 
and interpretation of the model and results. 
All authors contributed to the final writing and approved the manuscript.


\section*{Competing interests}

The authors declare no competing interests. 


\newpage

\renewcommand{\figurename}{\textbf{Supplementary Figure}}
\renewcommand{\tablename}{\textbf{Supplementary Table}}
\setcounter{figure}{0}
\setcounter{table}{0}


\begin{center}
	\Large{\bf Optimal control of the COVID-19 pandemic:\\ 
		controlled sanitary deconfinement in Portugal}
\end{center}

\bigskip
\bigskip

\begin{center}
\large C.~J.~Silva, C.~Cruz, D.~F.~M. Torres, A.~P.~Mu\~nuzuri, A.~Carballosa, I.~Area, 
J.~J.~Nieto, R.~Fonseca-Pinto, R.~Passadouro~da~Fonseca, E.~Soares~dos~Santos, W.~Abreu, J.~Mira\\[1.3cm]
\bf \Large --- Supplementary Information ---
\end{center}	

\newpage


\section*{}


\section*{Supplementary Figures}
\label{sec:supplement}

\begin{figure}[!htb]
\centering 
\includegraphics[scale=0.5]{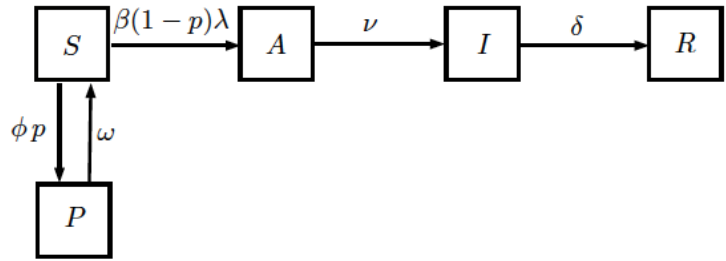}
\caption{\textbf{Diagram of the $\mathbf{SAIRP}$ model for the transmission dynamics 
of SARS-CoV-2 in a homogeneous population.} The population is subdivided into five compartments 
depending on the state of infection and disease of the individuals: $S$, susceptible 
(uninfected and not immune); $A$, infected but asymptomatic (undetected); $I$, 
active infected (symptomatic and detected/confirmed); $R$, removed 
(recovered and deaths by COVID-19); $P$, \emph{protected/prevented} 
(not infected, not immune, but that are under protective measures). }
\label{fig:diagrama}
\end{figure}

\begin{figure}[!htb]
\centering 
\subfloat[]{\label{fig:HospInf:realdata}
\includegraphics[scale=0.5]{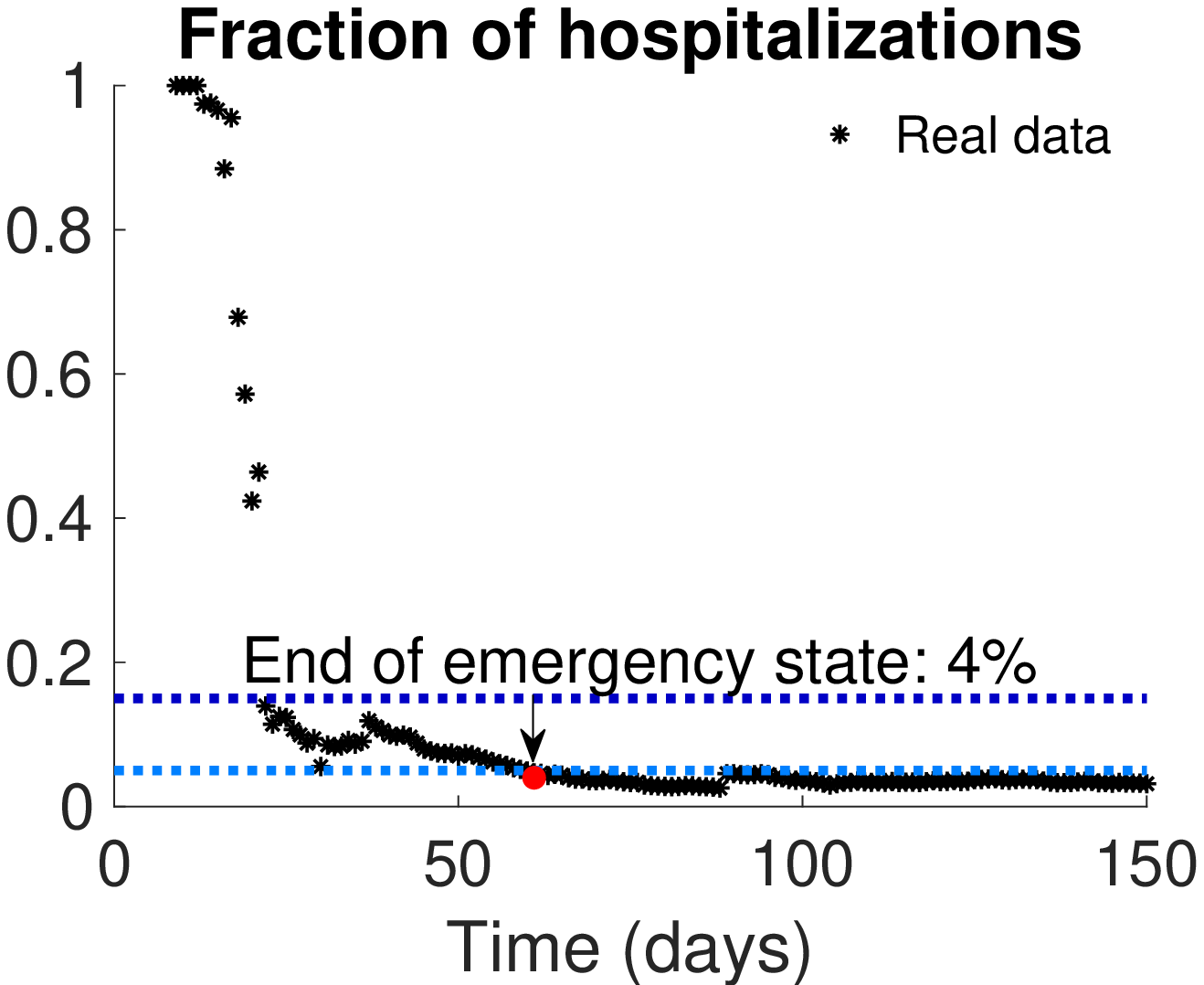}}
\subfloat[]{\label{fig:uciInf:realdata}
\includegraphics[scale=0.5]{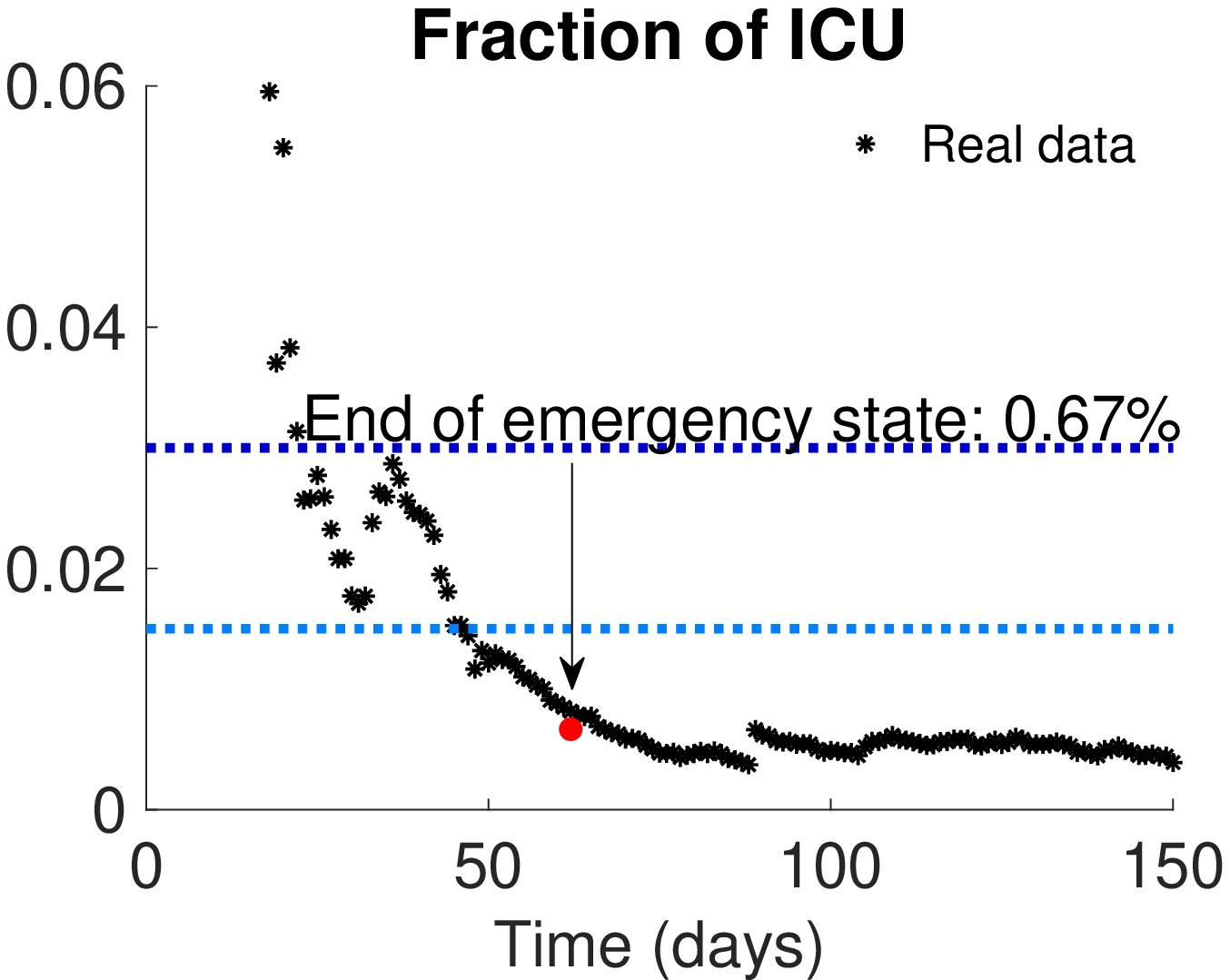}} 
\caption{ \textbf{Official real data, from March 02 to July 29, for the fraction 
of hospitalized individuals and in ICU due to COVID-19, 
with respect to the confirmed/active infected individuals.}    
(a) Fraction of hospitalized individuals due to COVID-19 
with respect to the number of active infected individuals, $H/I$.
(b) Fraction of intensive care units (ICU) hospitalized individuals 
due to COVID-19 with respect to the number of active infected individuals, $ICU/I$. }
\label{fig:HospICU:Inf:realdata}
\end{figure}

\begin{figure}[!htb]
\centering 
\subfloat[]{\label{fig:variar:m:1}
\includegraphics[width=0.5\textwidth]{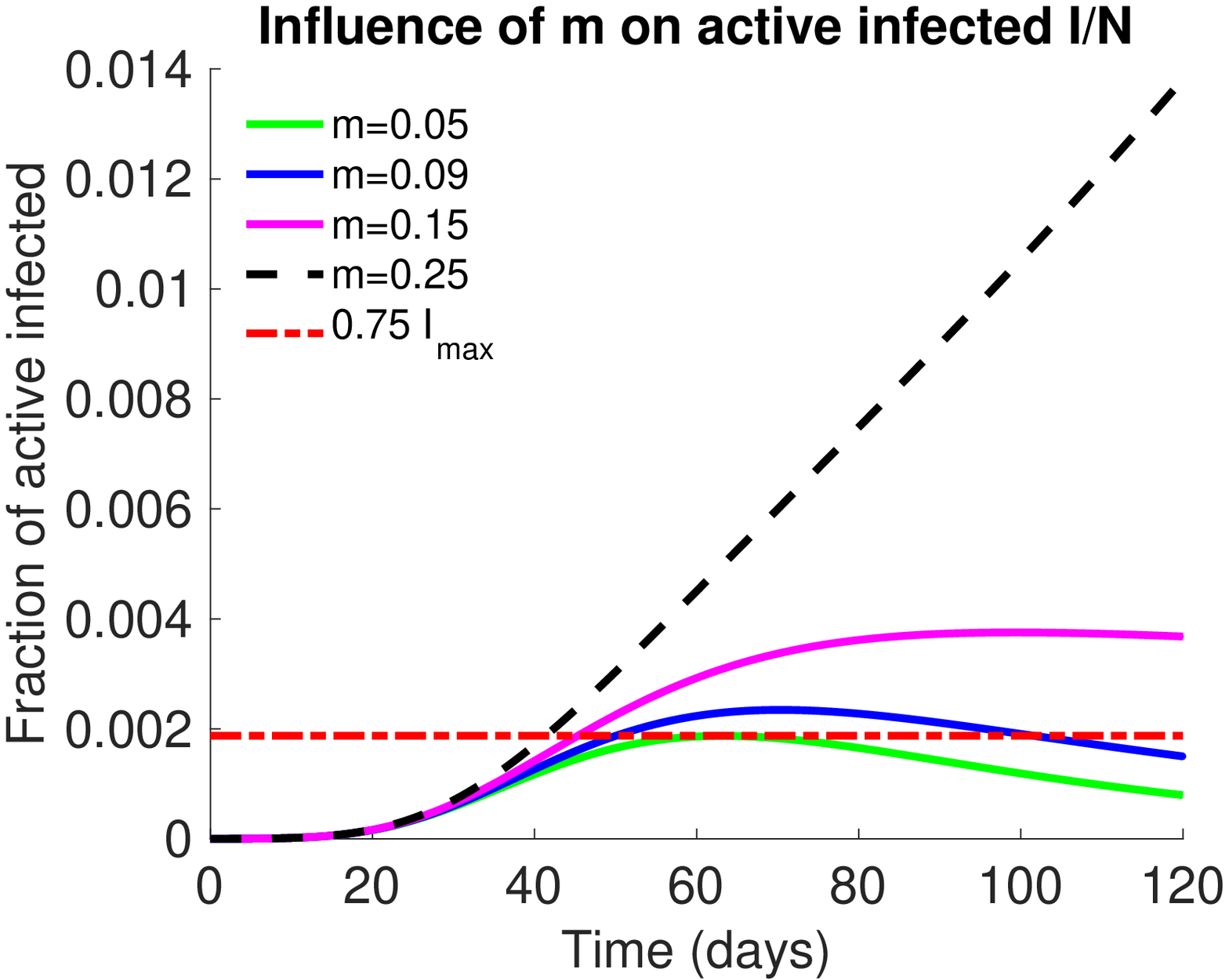}}
\subfloat[]{\label{fig:variar:m:2}
\includegraphics[width=0.5\textwidth]{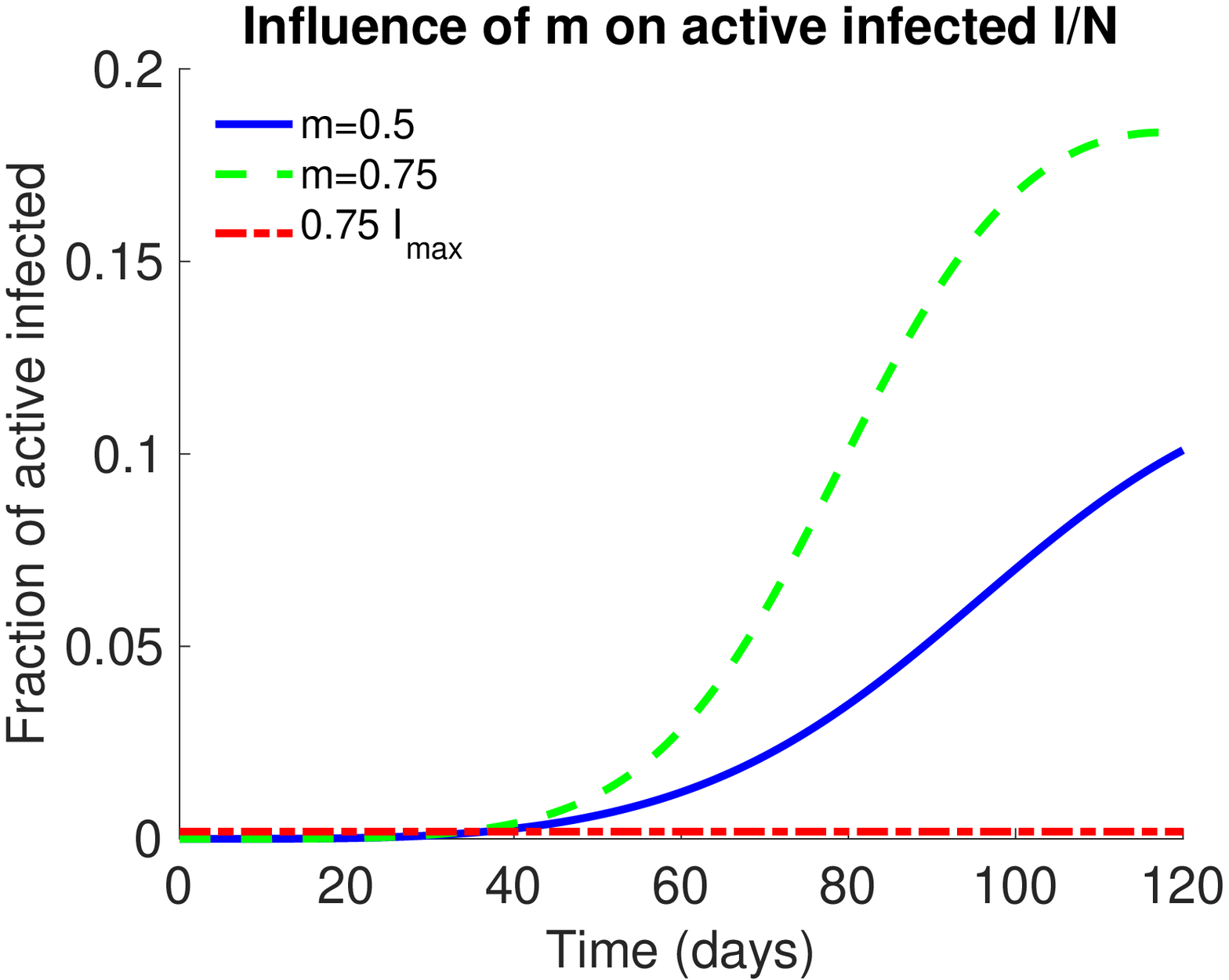}}
\caption{\textbf{Sensitivity of class $\mathbf{I}$ with respect to parameter $\mathbf{m}$.}  
Fraction of active infected individuals $I$ for: 
(a) $m \in \{ 0.05, 0.09, 0.15, 0.25 \}$, the dotted red line marks 
the level $0.75 \times I_{\max}$ that represents approximately 75\% 
of the maximum fraction of active infected cases observed in Portugal (up to July 29, 2020);
(b) $m \in \{ 0.5, 0.75 \}$, the dotted red line marks the level 
$0.75 \times I_{\max}$ that represents approximately 75\% 
of the maximum fraction of active infected cases observed in Portugal. 
We consider the fixed parameters $(\beta, p) = (1.464, 0.675)$ and all the other parameters
from Table~3 in Methods. }
\label{fig:variar:m}
\end{figure}

\begin{figure}[!htb]
\centering
\subfloat[]{\includegraphics[width=0.5\textwidth,height=50mm]{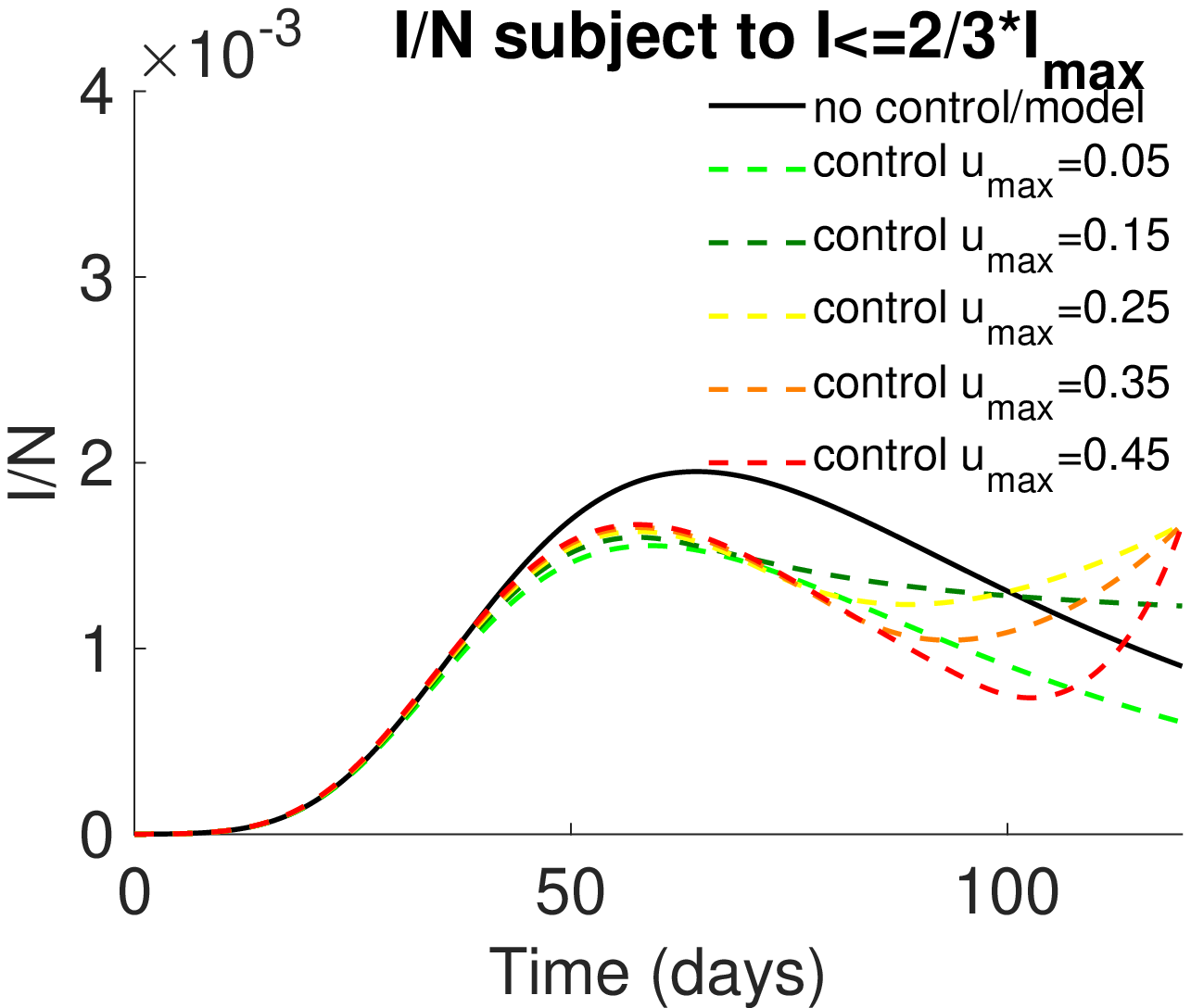} }
\subfloat[]{\includegraphics[width=0.5\textwidth, height=50mm]{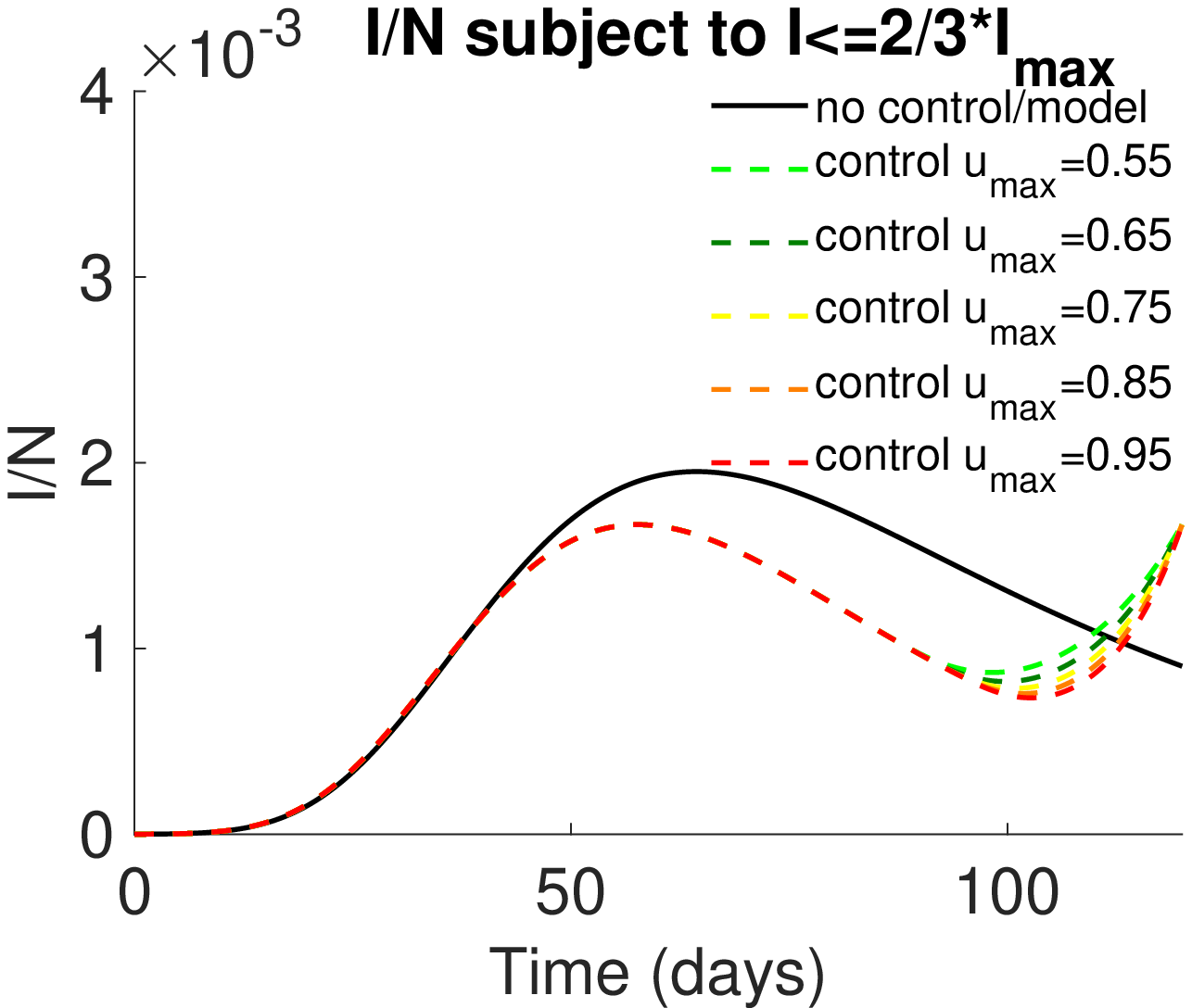} }
\caption{\textbf{Fraction of active infected individuals $\mathbf{I/N}$}. With control (dotted colored lines) 
and without control (continuous black line). The controlled solutions are subject the constraint 
$I(t)\leq \frac{2}{3} \times I_{\max}$ and different values of 
$u_{\max}=\{0.05, 0.15, 0.25, 0.35, 0.45, 0.55, 0.65, 0.75, 0.85, 0.95 \}$.}
\end{figure}

\begin{figure}[!htb]
\centering
\subfloat[]{\includegraphics[width=0.5\textwidth, height=50mm]{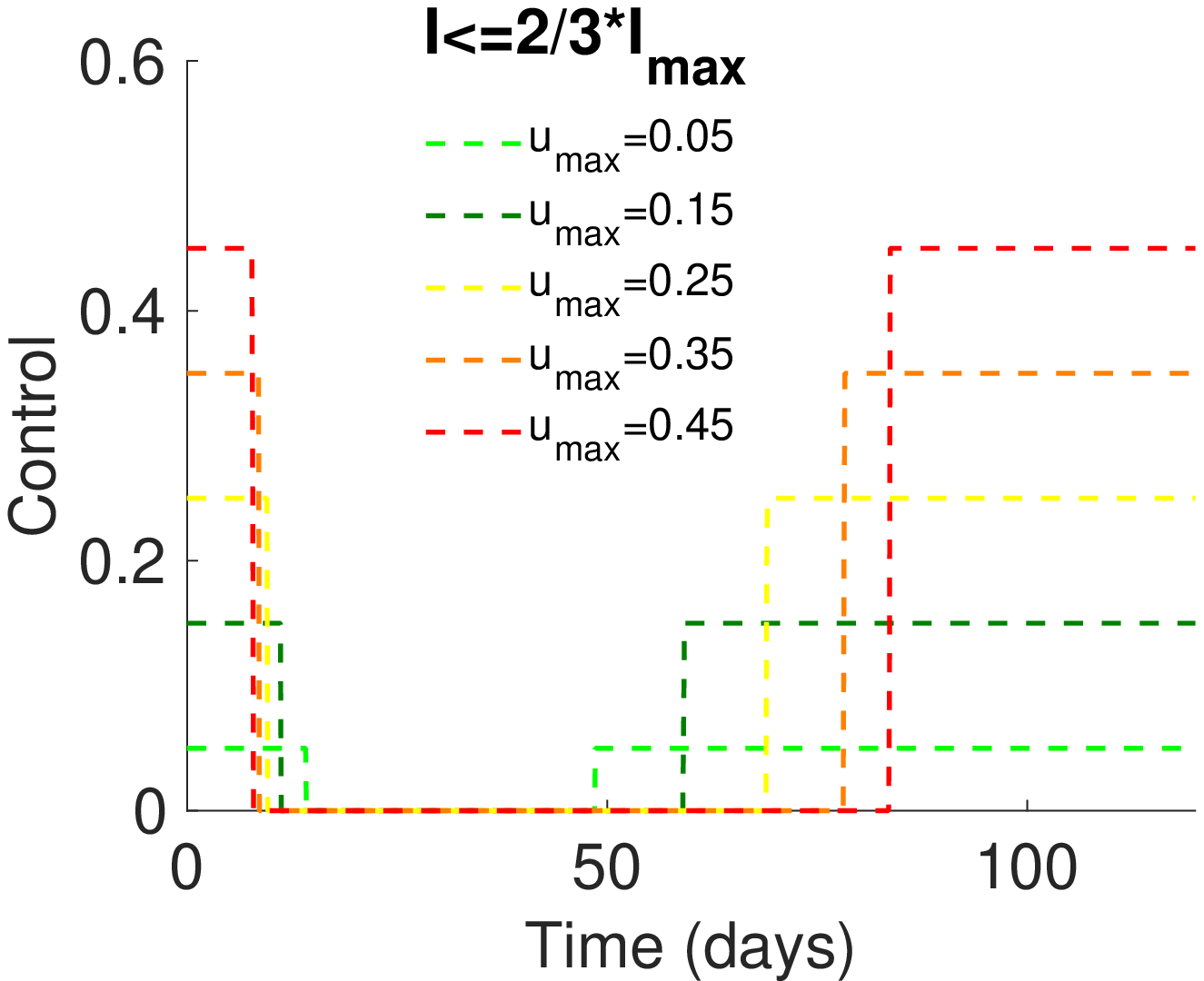}}
\subfloat[]{\includegraphics[width=0.5\textwidth, height=50mm]{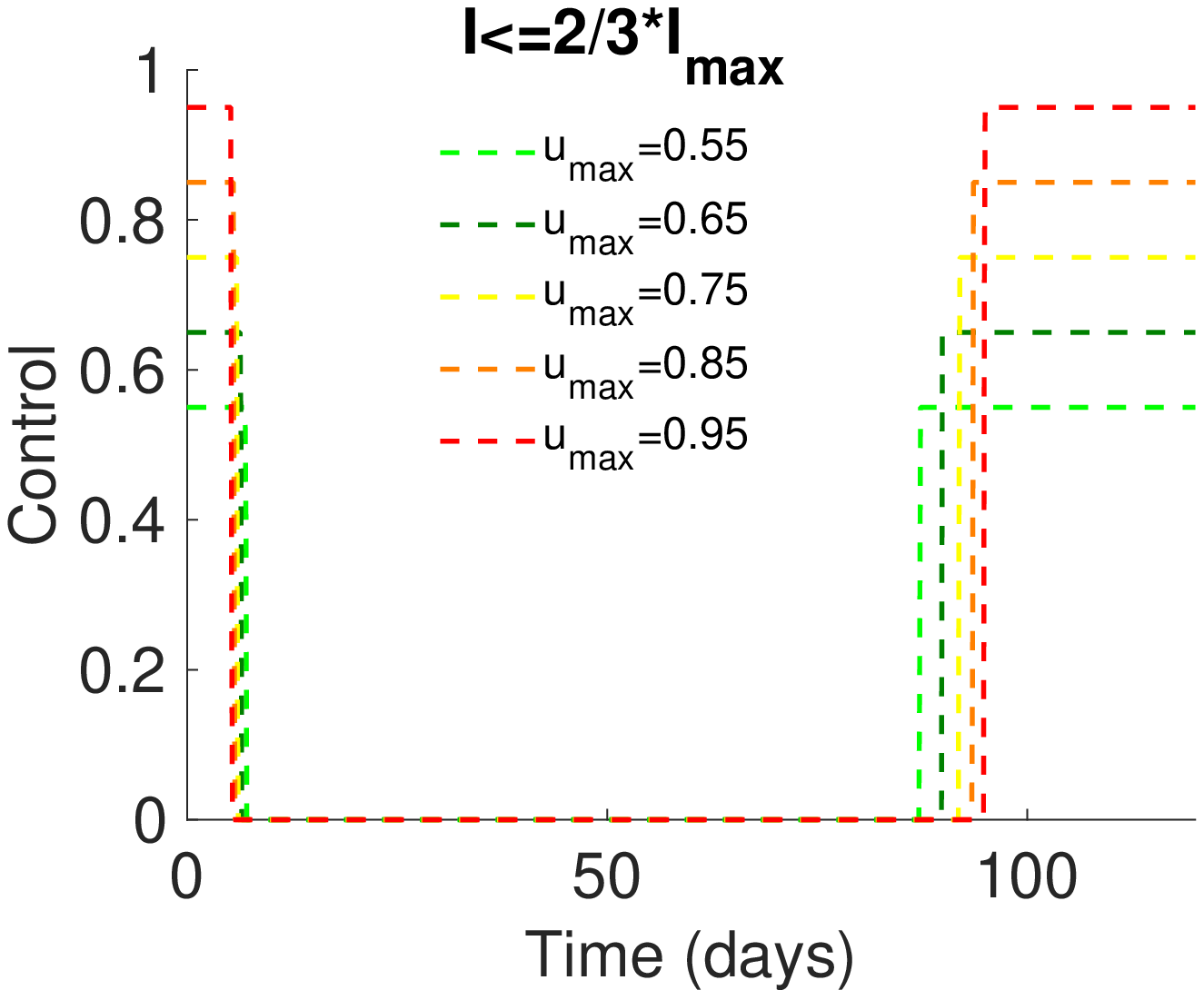}}
\caption{\textbf{Solution $\mathbf{u}$ of the optimal control problem.} 
Considering the state constraint $I(t)\leq \frac{2}{3} \times I_{\max}$ 
and different values of $u_{\max}=\{0.05, 0.15, 0.25, 0.35, 0.45, 0.55, 0.65, 0.75, 0.85, 0.95 \}$.}
\end{figure}

\begin{table}[!htb]
\caption{\textbf{Analysis of the time interval with no transfer 
from $\mathbf{P}$ to $\mathbf{S}$  ($\mathbf{u_{\mathbf{\max}} =0}$)
after releasing $\mathbf{u_{\mathbf{\max}}}$ individuals in the first period.} 
The control takes the maximum value $u_{\max}$, considering the constraint 
$I(t)\leq \frac{2}{3} \times I_{\max}$.}
\label{TabelaGraficoPortugalControlo_1}
\centering
\begin{tabular}[center]{ l l | l l }  \hline
Control $u(\cdot)$ & Time interval & Control $u(\cdot)$ & Time interval  \\  \hline 
$u_{\max} = 0.05$ & $\approxeq 34.4 \text{ days}$ & $u_{\max} = 0.55$  & $\approxeq 79.8 \text{ days}$ \\
$\color{green}{u_{\max} = 0.10}$ &  $\approxeq 42.6 \text{ days}$ & $u_{\max} = 0.60$  & $\approxeq 81.8 \text{ days}$ \\
$u_{\max} = 0.15$ & $\approxeq 47.9 \text{ days}$ & $u_{\max} = 0.65$  & $\approxeq 83.3 \text{ days}$ \\
$\color{OliveGreen}{u_{\max} = 0.20}$ & $\approxeq 52.1 \text{ days}$ & $u_{\max} = 0.70$  & $\approxeq 84.6 \text{ days}$ \\
$u_{\max} = 0.25$ & $\approxeq 59.4 \text{ days}$ & $u_{\max} = 0.75$  & $\approxeq 85.8 \text{ days}$ \\
$\color{yellow}{u_{\max} = 0.30}$ & $\approxeq 65.3 \text{ days}$ & $u_{\max} = 0.80$   & $\approxeq 86.8 \text{ days}$ \\
$u_{\max} = 0.35$ & $\approxeq 69.6 \text{ days}$ & $u_{\max} = 0.85$ & $\approxeq 87.8 \text{ days}$ \\
$\color{Orange}{u_{\max} = 0.40}$ & $\approxeq 73 \text{ days}$ &  $u_{\max} = 0.90$ & $\approxeq 88.6 \text{ days}$ \\
$u_{\max} = 0.45$ & $\approxeq 75.7 \text{ days}$ & $u_{\max} = 0.95$  &  $\approxeq 89.5 \text{ days}$\\
$\color{red}{u_{\max} = 0.50}$  & $\approxeq 78.1 \text{ days}$  &   &   \\	 \hline 
\end{tabular}
\end{table}

\begin{figure}[!htb]
\centering
\includegraphics[scale=0.6]{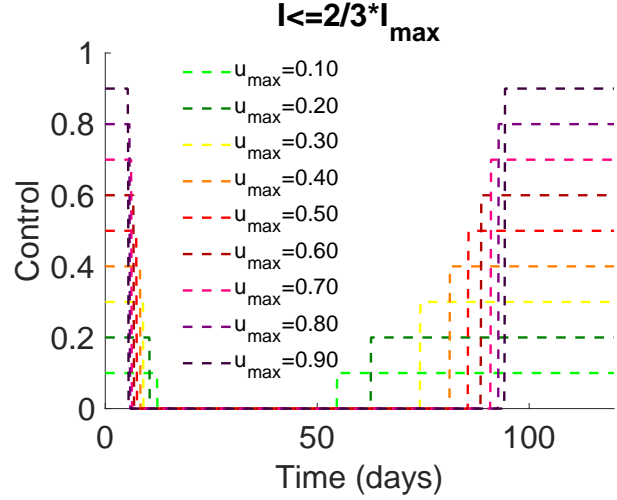}
\caption{\textbf{Solution of the optimal control problem $\mathbf{u}$ 
subject to the state constraint $\mathbf{I(t)\leq \frac{2}{3} \times I_{\max}}$ 
and different values of 
$\mathbf{u_{\max} = \{ 0.10, 0.20, 0.30, 0.40, 0.50, 0.60, 0.70, 0.80, 0.90  \}}$.}}
\end{figure}

\begin{figure}[!htb]
\centering
\subfloat[]{\includegraphics[scale=0.5]{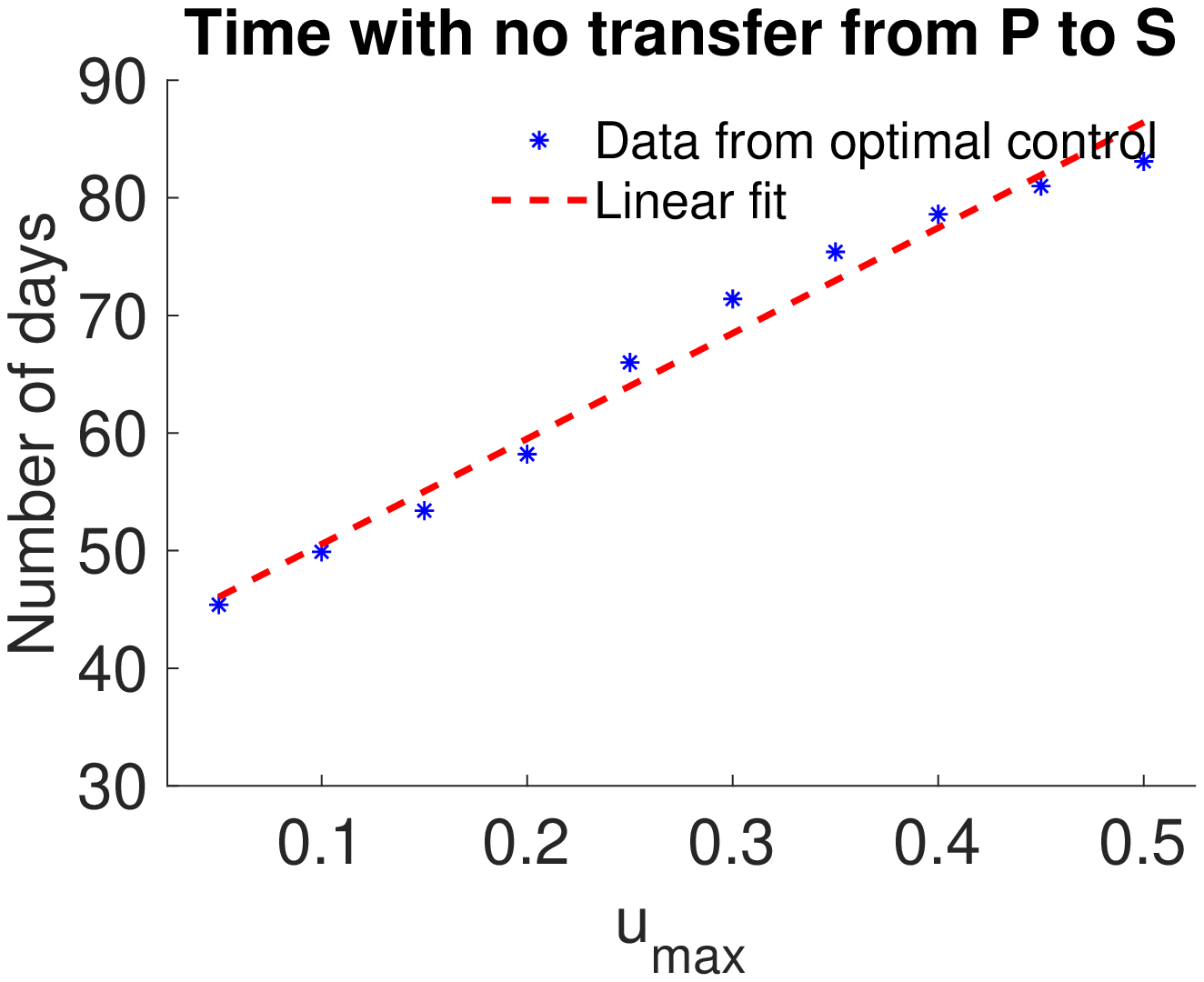}}
\subfloat[]{\includegraphics[scale=0.5]{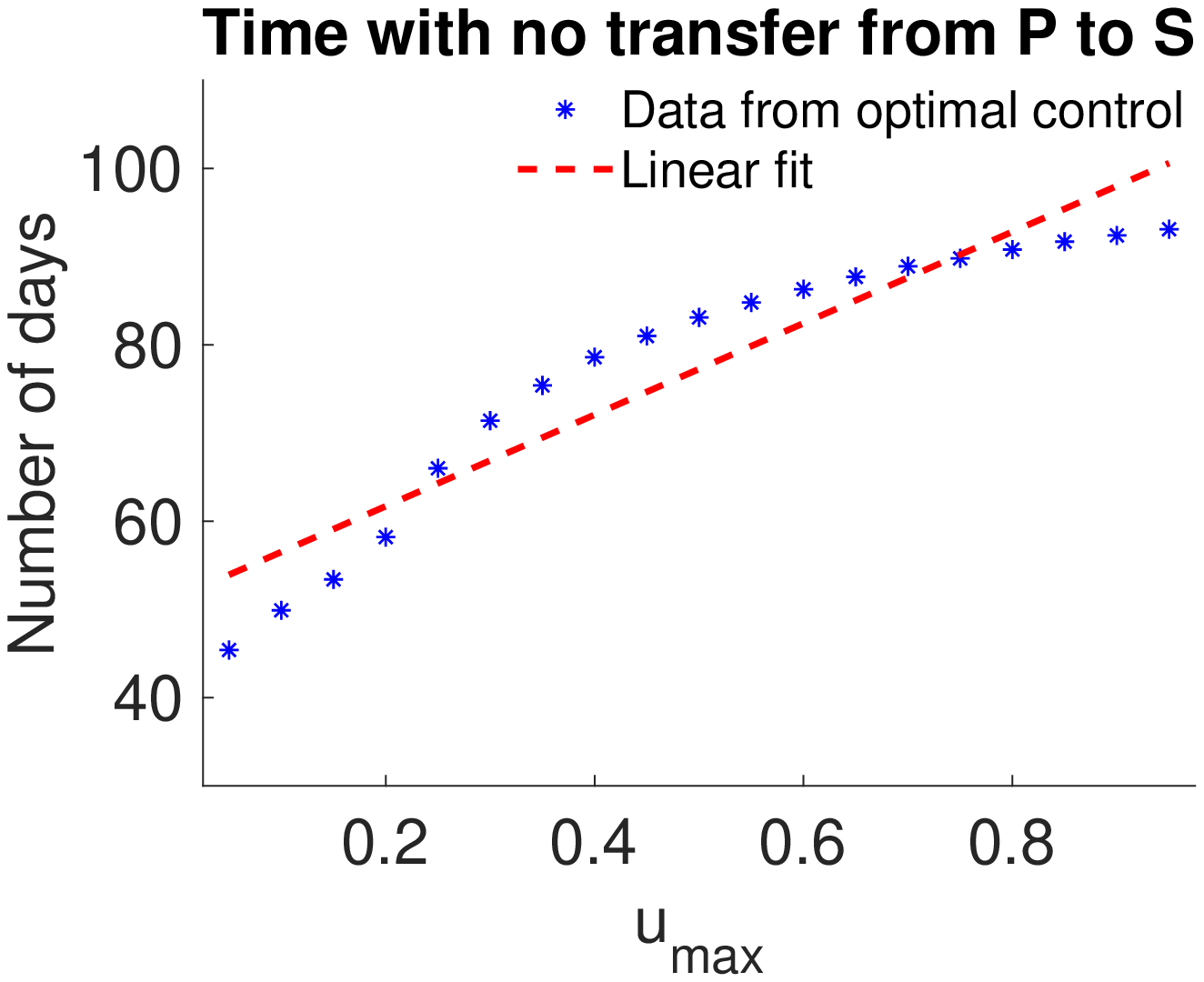}}
\caption{\textbf{Time with no transfer from $\mathbf{P}$ to $\mathbf{S}$ subject 
to $\mathbf{I \leq 0.60 \times I_{max}}$ with a linear fit analysis.} 
Analysis of the relation/pattern between the maximal value $u_{\max}$ 
of the control and the number of days where there are no transfer of individuals from 
class $P$ to the class $S$, here referred as the \emph{no transfer time interval}, 
after having released the fraction $u_{\max}$ of persons from class $P$ to class $S$. 
In (a) the discontinuous red line is obtained by the linear fit $y = 89.697 x + 41.573$ 
for $u_1\in[0;0.50]$  and in (b) by $y = 51.863 x + 51.326$ for $u_1\in[0;0.95]$, 
where $y$ corresponds to the number of days with no transfer of individuals from $P$ to $S$, 
after having released the fraction $x$ (equiv. $u$) of class $P$ to class $S$ 
(the linear fit is obtained by means of a standard linear regression procedure).}
\label{FigControlLinearFit}
\end{figure}

\begin{figure}[!htb]
\centering
\includegraphics[scale=0.60]{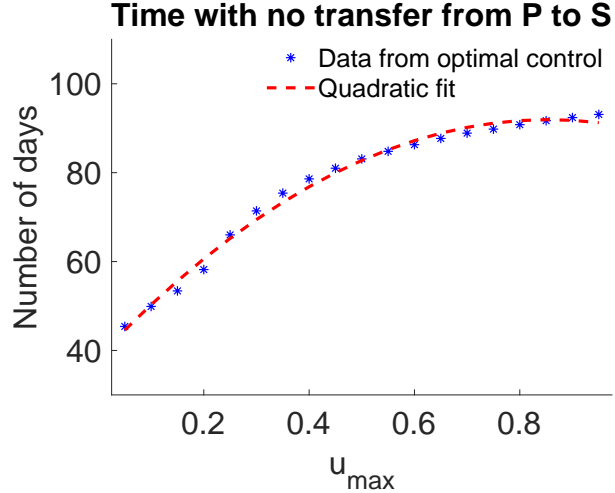}
\caption{\textbf{Time with no transfer from $\mathbf{P}$ to $\mathbf{S}$ 
subject to $\mathbf{I \leq 0.60 \times I_{\max}}$ with a quadratic fit analysis.} 
Analysis of the relation between the maximal value $u_{\max}$ of the control 
and the number of days that there are no transfer of individuals from 
class $P$ to the class $S$, considering a quadratic fit (red discontinuous line) 
$y = -73.251x^2+125.114x+38.507$ and $u_{\max}\in[0;0.95]$ 
w.r.t. time with no transfer from $P$ to $S$.}
\label{FigControlQuadraticFit}
\end{figure}

\begin{figure}[!htb]
\centering
\includegraphics[scale=0.45]{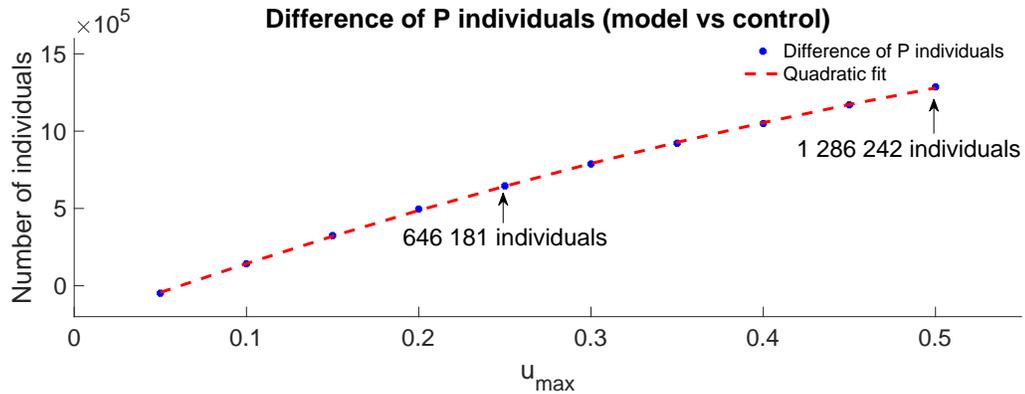}
\caption{\textbf{Difference between protected individuals obtained 
via the considered SAIRP model and the model with control}.  
Consider the maximal value of the control $u_{\max}\in\{0.05, 0.10, \dots, 0.45, 0.50\}$ and 
the constraint $I(t)\leq 0.60 \times I_{\max}$. 
The quadratic equation for fitting the difference between the number 
of individuals in class $P$ obtained via de $SAIRP$ model without 
and with control $u_{\max}\in\{0.05, 0.10, \dots, 0.45, 0.50\}$ 
(that is the number of released people from the protected class to the susceptible), 
respectively, is given by 
$y = -1984603.049 \, x^2 + 4030952.677 \, x  -239897.361$.}
\end{figure}

\begin{figure}[!htb]
\centering
\includegraphics[scale=0.60]{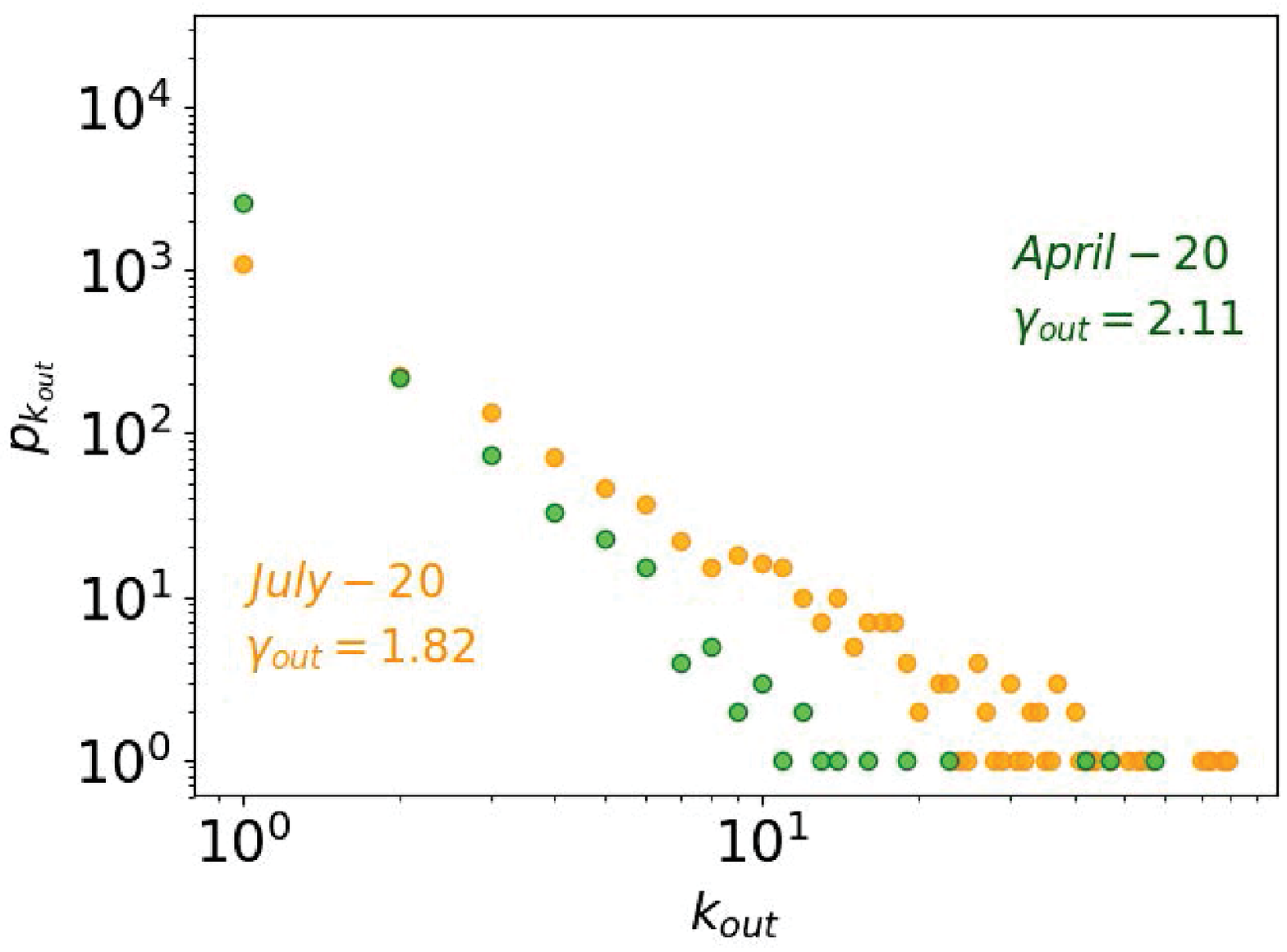}
\caption{\textbf{Connectivity distribution of the social networks obtained 
as described in the text.} Dots in green, as well as information in green, 
the network obtained in April 2020, while yellow dots correspond 
to the situation in July 2020. In both cases, the network topology corresponds 
to a scale free network with an exponent $\gamma=2.11$ in April 
and $\gamma=1.82$ in July 2020.}
\label{fig:methods:network}
\end{figure}


\begin{thebibliography}{10}
	
\bibitem{Noah}
\newblock Peeri, N. C. et al.
\newblock The SARS, MERS and novel coronavirus (COVID-19) epidemics, 
the newest and biggest global health threats: what lessons have we learned?
\newblock \emph{Int. J. Epidemiol.} \textbf{49}, 717--726 (2020).
	
\bibitem{worldometers}
\newblock COVID-19 Coronavirus Pandemic.
\newblock \url{https://www.worldometers.info/coronavirus/} (2020).

\bibitem{aulas:susp}
\newblock Rep\'ublica Portuguesa, Minist\'erio da Educa\c{c}\~ao, XXII Governo.
\newblock \emph{Comunica\c{c}\~ao enviada \`as escolas sobre suspens\~ao das atividades 
com alunos nas escolas de 16 de mar\c{c}o a 13 de abril}.
\newblock \url{https://www.portugal.gov.pt/pt/gc22/comunicacao/documento?i=comunicacao-enviada-as-escolas-sobre-suspensao-das-atividades-com-alunos-nas-escolas-de-16-de-marco-a-13-de-abril}
(2020).
	
\bibitem{cap:UCI:pt}
\newblock \emph{Capacidade de Medicina Intensiva aumentou 23\%}.
\newblock \url{https://covid19.min-saude.pt/capacidade-de-medicina-intensiva-aumentou-23/}
(2020).
	
\bibitem{Metcalf}
\newblock Metcalf, C. J. E. , Morris, D. H. \& Park, S. W.
\newblock Mathematical models to guide pandemic response.
\newblock \emph{Science} \textbf{369}, 368--369 (2020).
	
\bibitem{Giordano:modcovid:NatMed}
\newblock Giordano, G. et al.
\newblock Modelling the COVID-19 epidemic and implementation of population-wide interventions in Italy.
\newblock \emph{Nat. Med.} \textbf{26}, 855--860 (2020).
	
\bibitem{Lopez:HumBehav2020}
\newblock L\'{o}pez, L. \& Rod\'{o}, X.
\newblock The end of social confinement and COVID-19 re-emergence risk.
\newblock \emph{Nat. Hum. Behav.} \textbf{4}, 746--755 (2020).

\bibitem{Hoertel:modFrance:Natmed}
\newblock Hoertel, N. et al.
\newblock A stochastic agent-based model of the SARS-CoV-2 epidemic in France.
\newblock \emph{Nat. Med.} (2020).

\bibitem{Kissler860}
\newblock Kissler, S. M., Tedijanto, C., Goldstein, E., Grad, Y. H. \& Lipsitch, M.
\newblock Projecting the transmission dynamics of SARS-CoV-2 through the postpandemic period.
\newblock  \emph{Science} \textbf{368}, 860--868 (2020).

\bibitem{MR4091761}
\newblock Campos, C., Silva, C. J. \& Torres, D. F. M.  
\newblock Numerical optimal control of HIV transmission in Octave/MATLAB. 
\newblock \emph{Math. Comput. Appl.} {\bf 25}, no.~1, 20~pp (2020). 
\newblock {\tt arXiv:1912.09510}

\bibitem{MR3918295}
\newblock Malinzi, J., Ouifki, R., Eladdadi, A., Torres, D. F. M. \& White, K. A. J.  
\newblock Enhancement of chemotherapy using oncolytic virotherapy: 
mathematical and optimal control analysis. 
\newblock \emph{Math. Biosci. Eng.} {\bf 15}, no.~6, 1435--1463 (2018). 
\newblock {\tt arXiv:1807.04329}

\bibitem{MR3629459}
\newblock Sharomi, O. \& Malik, T.  
\newblock Optimal control in epidemiology. 
\newblock \emph{Ann. Oper. Res.} {\bf 251}, no.~1-2, 55--71  (2017). 

\bibitem{Rawson:OC}
\newblock Rawson, T., Brewer, T., Veltcheva, D., Huntingford, C. \& Bonsall, M. B.
\newblock How and When to End the COVID-19 Lockdown: An Optimization Approach.
\newblock \emph{Front. Public Health} \textbf{8}, 262 (2020).
	
\bibitem{Tsay:OC:US}
\newblock Tsay, C. et al.
\newblock Modeling, state estimation, and optimal control for the US COVID-19 outbreak.
\newblock \emph{Sci. Rep.} \textbf{10}, 10711 (2020).

\bibitem{Libotte}
\newblock Libotte, G. B., Lobato, F. S., Platt, G. M. \& Neto, A. J. S.
\newblock Determination of an optimal control strategy 
for vaccine administration in COVID-19 pandemic treatment.
\newblock \emph{Comput. Methods Programs Biomed.} \textbf{196}, 105664 (2020).

\bibitem{Lemecha}
\newblock Obsu, L. L. \& Balcha, S. F.
\newblock Optimal control strategies for the transmission risk of COVID-19.
\newblock \emph{J. Biol. Dyn.} \textbf{14}, 590--607 (2020).

\bibitem{Zine}
\newblock Zine, H., Boukhouima, A., Lotfi, E. M., Mahrouf,  M., Torres, D. F. M.  \& Yousfi, N. 
\newblock A stochastic time-delayed model for the effectiveness 
of Moroccan COVID-19 deconfinement strategy.
\newblock \emph{Math. Model. Nat. Phenom.} \textbf{15}, Art. 50, 14~pp (2020). 
\newblock {\tt arXiv:2010.16265}

\bibitem{Moradian:2020}
\newblock Moradian, N. et al.
\newblock The urgent need for integrated science to fight COVID-19 pandemic and beyond.
\newblock \emph{J. Transl. Med.} \textbf{18}, 205 (2020).
	
\bibitem{protective:Lancet}
\newblock Chu, D. K. et al.
\newblock Physical distancing, face masks, and eye protection to prevent 
person-to-person transmission of SARS-CoV-2 and COVID-19: a systematic review and meta-analysis.
\newblock \emph{The Lancet} \textbf{395}, 1973--1987 (2020).

\bibitem{Haug:NatHB2020}
\newblock Haug, N. et al. 
\newblock Ranking the effectiveness of worldwide COVID-19 government interventions. 
\newblock \emph{Nat Hum Behav} (2020). 
	
\bibitem{dgs-covid}
\newblock {Dire\c{c}\~{a}o-Geral da Sa\'{u}de -- COVID-19},
\newblock {Ponto de Situa\c{c}\~{a}o Atual em Portugal}.
\newblock \url{https://covid19.min-saude.pt/ponto-de-situacao-atual-em-portugal/} (2020).
	
\bibitem{legislacao:covid19}
\newblock Legisla\c{c}\~{a}o Compilada -- COVID-19.
\newblock \url{https://dre.pt/legislacao-covid-19-upo} (2020).

\bibitem{Watts1998}
\newblock Watts, D.J. \& Strogatz, S.H.
\newblock Collective dynamics of small-world networks
\newblock \emph{Nature} \textbf{393}, 440--442 (1998).

\bibitem{Teslya:PLOSMed2020}
\newblock Teslya, A. et al.
\newblock Impact of self-imposed prevention measures and short-term 
government-imposed social distancing on mitigating and delaying 
a COVID-19 epidemic: A modelling study.
\newblock \emph{PLoS Med.} \textbf{17}(7):e1003166 (2020).

\bibitem{Moghadas}
\newblock Moghadas, S. M. et al.
\newblock The implications of silent transmission for the control of COVID-19 outbreaks.
\newblock \emph{Proc. Natl. Acad. Sci. U.S.A.} \textbf{117}, 17513--17515 (2020).

\bibitem{Driessche}
\newblock Driessche, P. van~den \& Watmough, J.
\newblock Reproduction numbers and sub-threshold endemic equilibria 
for compartmental models of disease transmission.
\newblock \emph{Math. Biosci.} \textbf{180}, 29--48 (2002).

\bibitem{cdc:param:est}
\newblock COVID-19 Pandemic Planning Scenarios.
\newblock \url{https://www.cdc.gov/coronavirus/2019-ncov/hcp/planning-scenarios.html} (2020).

\bibitem{Li:Science:2020} 
\newblock Li, R. et al.
\newblock Substantial undocumented infection facilitates 
the rapid dissemination of novel coronavirus (SARS-CoV-2).
\newblock \emph{Science} \textbf{368}, 489--493 (2020).

\bibitem{Mizumoto:2019}
\newblock Mizumoto, K., Kagaya, K., Zarebski, A. \& Chowell, G.
\newblock Estimating the asymptomatic proportion of coronavirus 
disease 2019 (COVID-19) cases on board the Diamond Princess cruise ship, Yokohama, Japan, 2020.
\newblock \emph{Euro Surveill.} \textbf{25}(10):2000180 (2020).

\bibitem{Park:Epidemics:2020}
\newblock Park, S. W., Cornforth, D. M., Dushoff, J. \& Weitz, J. S.
\newblock The time scale of asymptomatic transmission affects estimates 
of epidemic potential in the COVID-19 outbreak.
\newblock \emph{Epidemics} \textbf{31}:100392 (2020).

\bibitem{lancet:timerecover:2020}
\newblock Bi, Q. et al.
\newblock Epidemiology and transmission of COVID-19 in 391 cases 
and 1286 of their close contacts in Shenzhen, China: a retrospective cohort study.
\newblock \emph{Lancet Infect. Dis.} \textbf{20}, 911--919 (2020).

\bibitem{AnaPaiao:Ecology2020}
\newblock Lemos-Pai\~ao A. P., Silva, C. J. \& Torres, D. F. M.
\newblock A new compartmental epidemiological model for COVID-19 with a case study of Portugal.
\newblock \emph{Ecol. Complex.} \textbf{44}, 100885 (2020).
\newblock {\tt arXiv:2011.08741}

\bibitem{python}
\newblock Python package GetOldTweets3\\
\newblock \url{https://pypi.org/project/GetOldTweets3/}

\bibitem{INE}
\newblock Statistics Portugal,
\newblock \url{https://www.ine.pt/xportal/xmain?xpid=INE&xpgid=ine_indicadores&contecto=pi&indOcorrCod=0008273&selTab=tab0}
(2020). 

\bibitem{Albert2002}
\newblock Albert, R.,  Barabasi, A.-L.
\newblock Statistical mechanics of complex networks.
\newblock \emph{Rev. Mod. Phys.} \textbf{74}, 47 (2002).

\bibitem{Pereira2016}
\newblock Pereira, F. S., de Amo, S. \& Gama, J.
\newblock Evolving centralities in temporal graphs: a twitter network analysis.
\newblock \emph{17th IEEE International Conference on Mobile Data Management (MDM)} 
\textbf{2}, 43--48 (2016).

\bibitem{Abel2011}
\newblock Abel, F., Gao, Q., Houben, G. J. \& Tao, K.
\newblock Analyzing temporal dynamics in twitter profiles 
for personalized recommendations in the social web.
\newblock \emph{Proceedings of the 3rd International Web Science Conference}, 1--8 (2011).

\bibitem{Cataldi2010}
\newblock Cataldi, M., Di Caro, L. \& Schifanella, C.
\newblock Emerging topic detection on twitter based on temporal and social terms evaluation.
\newblock \emph{Proceedings of the tenth international workshop on multimedia data mining}, 1--10 (2010).

\bibitem{Verhulst1845}
\newblock Verhulst, P. F.
\newblock Resherches mathematiques sur la loi d'accroissement de la population.
\newblock \emph{Nouveaux memoires de l'academie royale des sciences} \textbf{18}, 1--41 (1845) 
(in French).
	
\bibitem{Lloyd1995}
\newblock Lloyd, A. L.
\newblock The coupled logistic map: a simple model for the effects 
of spatial heterogeneity on population dynamics.
\newblock \emph{J. Theor. Biol.} \textbf{173}, 217--230 (1995).
	
\bibitem{Tarasova2017}
\newblock Tarasova, V. V. \& Tarasov, V. E.
\newblock Logistic map with memory from economic model.
\newblock \emph{Chaos, Solitons \& Fractals} \textbf{95}, 84--91 (2017).
		
\bibitem{Carballosa2020}
\newblock Carballosa, A., Mussa-Juane, M. \& Mu\~{n}uzuri, A.P.
\newblock Incorporating social opinion in the evolution of an epidemic spread.
\newblock Submitted (2020). 
Preprint at \url{http://arxiv.org/abs/2007.04619}
	
\bibitem{AMPL}
\newblock Fourer, R., Gay, D. M. \& Kernighan, B. W.
\newblock \emph{AMPL: A Modeling Language for Mathematical Programming}.
\newblock Duxbury Press, Brooks–Cole Publishing Company (1993).
	
\bibitem{IPOPT}
\newblock W\"{a}chter, A. \& Biegler, L. T.
\newblock On the implementation of an interior-point filter line-search
algorithm for large-scale nonlinear programming.
\newblock \emph{Math. Program.} \textbf{106}, 25--57 (2006).
	
\bibitem{SilvaMaurerTorres}
\newblock Silva, C. J., Maurer, H. \& Torres, D. F. M.
\newblock Optimal control of a tuberculosis model with state and control delays.
\newblock \emph{Math. Biosci. Eng.} \textbf{14}, 321--337 (2017).
\newblock {\tt arXiv:1606.08721}

\end{thebibliography}
\end{document}